\documentclass[11pt,a4paper]{amsart}

\pdfoutput=1

\usepackage{booktabs}
\usepackage{lscape}%to display tables vertically

\usepackage{epstopdf}% To incorporate .eps illustrations using PDFLaTeX, etc.
\usepackage{subfigure}% Support for small, `sub' figures and tables
\usepackage{xcolor}% for different text colors
\usepackage{natbib}% Citation support using natbib.sty
\usepackage{graphicx}% For \resizebox
\usepackage{multirow}%for multirow in a table
\bibpunct[, ]{(}{)}{,}{a}{}{,}% Citation support using natbib.sty
\theoremstyle{plain}% Theorem-like structures

\theoremstyle{definition}

\theoremstyle{remark}

\renewcommand{\Pr}{\mathsf{P}}

\begin{document}

\title[Estimation of Contextual Exposure]{Estimation of Contextual Exposure to HIV from GPS Data}\thanks{CONTACT A. Dobra. Email: adobra@uw.edu}

\author[Wu et al.]{Haoyang Wu\textsuperscript{a}, Zhaoxing Wu\textsuperscript{a}, Thulile Mathenjwa\textsuperscript{b}, Elphas Okango\textsuperscript{b}, Khai Hoan Tram\textsuperscript{c}, Margot Otto\textsuperscript{b}, Maxime Inghels\textsuperscript{d}, Paul Mee\textsuperscript{d}, Diego Cuadros\textsuperscript{e}, Hae-Young Kim\textsuperscript{f}, Till B\"{a}rnighausen\textsuperscript{b,g}, Frank Tanser\textsuperscript{b,h} and Adrian Dobra\textsuperscript{a}}

\address{\textsuperscript{a}Department of Statistics, University of Washington, Seattle, WA, USA; \textsuperscript{b}Africa Health Research Institute, Durban, South Africa;  \textsuperscript{c}Department of Medicine, University of Washington, Seattle, WA, USA; \textsuperscript{d} Lincoln Institute for Rural and Coastal Health, University of Lincoln, Lincoln, U.K.; \textsuperscript{e}Digital Epidemiology Laboratory, University of Cincinnati, Cincinnati, Ohio, USA; \textsuperscript{f} Department of Population Health, New York University Grossman School of Medicine, New York, NY, USA; \textsuperscript{g} Heidelberg Institute of Global Health, Heidelberg University, Heidelberg, Germany; \textsuperscript{h} South African Centre for Epidemiological Modelling and Analysis (SACEMA), School of Data Science and Computational Thinking, Stellenbosch University, Stellenbosch, South Africa}

\begin{abstract}
We present a comprehensive statistical methodological framework for estimating contextual exposure to HIV that includes local (grid-cell level) estimation of HIV prevalence and human activity space estimation based on GPS data. The development of our framework was necessary to analyze HIV surveillance and sociodemographic survey data in conjunction with GPS data collected in rural KwaZulu-Natal, South Africa, to study the mobility patterns of young people. Based on mobility and contextual exposure measures, we examine whether the sex and age of study participants systematically influence the extent and structure of their mobility patterns. We discuss techniques for investigating how the study participants' contextual exposure to HIV changes as their activity spaces expand beyond residential locations, as well as methods for identifying study participants who may be at increased risk of acquiring HIV.
KEYWORDS: Contextual HIV exposure; GPS-based mobility analysis; Activity space; HIV prevalence mapping
\end{abstract}

\maketitle

\date{\today} 

\tableofcontents

\section{Introduction} \label{sec:introduction}

We focus on the problem of estimating contextual exposure to HIV using GPS data. GPS-based estimation involves analyzing location data over time to assess individuals' exposure to conditions that may negatively impact their health and well-being. Such conditions may include environmental factors (e.g., pollution, radiation, noise, and weather) as well as social risk factors (e.g., crime, noise levels, infectious diseases, access to care, and social isolation). GPS data reveal time-activity patterns that indicate the amount of time spent in various locations, such as home, work, school, and parks. This approach identifies the contexts in which exposure to specific environmental and social risk factors occurs. Contextual exposure provides a precise measure of exposure, in contrast to traditional methods that depend on administrative boundaries (e.g., census tracts) or fixed locations (e.g., home addresses), which may not accurately represent the actual time spent at various locations\citep{marquet-et-2022}. Activity spaces, which are areas where people spend time or travel during their daily activities \citep{gesler-meade-1988}, can provide a more accurate measure of contextual risks \citep{kwan-2012}.  

Researchers track the movement patterns of study participants to assess their activity spaces and correlate these with contextual information. Geographic Information System (GIS) software integrates GPS data with various layers of environmental data \citep{clark-et-2025}. These layers can include air pollution levels, green spaces, noise levels, and proximity to roads, as well as environmental hazards, health resources, and social environments \citep{chaix-et-2013,yi-et-2019,marquet-et-2022}.  This framework offers insights into the variation of exposure in different locations and times, providing valuable information for health and social science research \citep{byrnes-et-2015,wiehe-et-2008}. Contextual exposure estimation using GPS data provides a more accurate perspective on exposure than static residential measures, facilitating the assessment of individual-specific exposure patterns. Furthermore, merging GPS data with high-resolution environmental information improves precision in exposure assessment\citep{marquet-et-2023}. 

Contextual exposure from GPS data has been employed in various types of studies, such as (a) air pollution studies: GPS data are key in understanding how different locations and activities contribute to individuals' exposure to air pollution \citep{wei-et-2025}; (b) exposure to green spaces:  GPS data are used to assess the amount of time spent in green spaces, potentially linking this with health outcomes \citep{clark-et-2025,marquet-et-2022}; (c) exposure to noise: GPS data can be combined with noise maps to study noise pollution levels in different areas \citep{zuo-et-2016}; and (d) studies of the built environment: GPS data can be integrated with data on walkability, access to amenities and other aspects of the built environment to understand their impact on health \citep{chaix-et-2013,glasgow-et-2019}.  

Current research utilizing GPS data in the context of HIV has mainly focused on mapping and characterizing areas with high levels of HIV transmission \citep{bulstra-et-2020}, as well as increasing participation in HIV care among individuals living with HIV \citep{hassani-et-2023}. Other key research areas include using GPS data to better understand how environmental factors contribute to HIV transmission \citep{duncan-et-2018,kandwal-et-2009}. Our contribution falls into this category. We utilize GPS data to define the activity spaces of an individual. We quantify individual contextual exposure based on estimates of HIV prevalence in spatial areas that encompass activity spaces. The fundamental premise is that people at increased risk for HIV acquisition are those with higher contextual exposures to HIV. This information is essential for public health officials to tailor interventions, such as targeted HIV testing and prevention programs, to specific locations where individuals at the highest risk spend a considerable portion of their time. 

The structure of this article is as follows. In Section \ref{sec:data}, we describe the HIV surveillance and sociodemographic data collected by the Africa Health Research Institute (AHRI) in rural KwaZulu-Natal, South Africa, and we introduce the Sesikhona GPS study that also took place in the AHRI study area. In Section \ref{sec:methods}, we describe our methodological framework, which comprises methods for longitudinal imputation of HIV status, determining local (grid cell level) estimates of HIV prevalence, estimating activity spaces from GPS data, and assessing contextual exposure to HIV. In Section \ref{sec:analysis} we present our analysis of the Sesikhona GPS data. We study the relationship between mobility and the demographic characteristics (sex and age) of the study participants, and we show how to combine GPS-based mobility measures with HIV contextual exposure to identify individuals at high risk of HIV acquisition. In Section \ref{sec:conclusions} we discuss the relevance and limitations of our methodology and our empirical results.

\section{Data} \label{sec:data}

We make use of two key sources of data, both collected at the Africa Health Research Institute (AHRI) in rural KwaZulu-Natal, South Africa: a population-based HIV surveillance system discussed in Section \ref{sec:hivsurveillance}, and a GPS dataset presented in Section \ref{sec:sesikhonastudy}.

\subsection{HIV Surveillance Data} \label{sec:hivsurveillance}

We utilized data from a major HIV cohort at the Africa Health Research Institute (AHRI) in rural KwaZulu-Natal, South Africa, covering the years 2011-2023. This country has the highest global HIV burden, with approximately 7.7 million people living with the virus \citep{UNAIDS2024report}. The KwaZulu-Natal region has historically reported some of the highest rates of HIV prevalence and incidence in the country \citep{birdthistle-et-2019}. The local community comprises approximately 140,000 individuals and is characterized by frequent migration, low marriage rates, polygamous marriages, multiple sexual partnerships, and limited knowledge and disclosure of HIV status \citep{dobra-et-2017}. The median time for men in the study cohort is 2,391 days (IQR = 4,549), while the median time for women is 2,384 (IQR = 4,541). Established in 2000, the Africa Center Demographic Information System (ACDIS), which is now part of AHRI, is a population-based surveillance system that covers approximately 140,000 people \citep{gareta-et-2021}. AHRI collects data on the characteristics of households and individuals that belong to family units in the rural study community. Individuals become part of the HIV cohort when they reach 15 years of age or immigrate to the rural study community \citep{tanser-et-2008}. The sampling was carried out over 12 demographic surveillance rounds (DSRounds), each identified by the ending year and lasting approximately 12 months. Between May 2018 and March 2020, the Vukuzazi multimorbidity survey and HDSS studies were conducted concurrently in the HDSS area. In 2019, finger prick samples were not collected for HDSS \citep{wong-et-2021}. Before 2020, each survey round was conducted from January to December. Since 2021, the timing has shifted to mid-year cycles due to disruptions caused by COVID-19. One notable exception is the DSRound that started in January 2020. This round was terminated in March and resumed in April 2021. It was completed in April 2022.

AHRI collects data on households and individuals in family units within the rural study community, regardless of their residency status. Births, deaths, and migrations are recorded every four months, while socioeconomic status is assessed annually. The residential locations within the AHRI study area have been accurately geolocated with an accuracy of $<$2m \citep{tanser-et-2009}. Study participants can change their place of residence multiple times. they may move between two residences located within the AHRI study area, between two residences located outside the AHRI study area, or between a residence inside the AHRI study area and another residence outside the AHRI study area. For individuals residing outside the community for a specific period, approximate locations are identified by place names gathered by field workers during family interviews. External migration events are concentrated in the metropolitan areas of Richards Bay, Durban, Johannesburg, and Pretoria \citep{dobra-et-2017}. The relevance of examining whether the study participants have resided outside the rural study area is derived from the findings of \citet{dobra-et-2017}. Their results indicate that, for the same rural study area, the risk of HIV acquisition is significantly increased for both men and women when they spend more time outside of this rural study area or when they change their residences over longer distances.  

\subsection{The Sesikhona GPS Study} \label{sec:sesikhonastudy}

The Sesikhona GPS study \citep{mathenjwa-et-2025} was conducted from June 2021 to May 2025 in the AHRI study area. The data were collected using Avicenna, a custom-built software that leverages Android location services to record the smartphone's location and upload the data to a secure, encrypted study server. A total of 207 participants were enrolled in three phases. Phase I involved 163 participants, while Phase II involved 44 participants. Phase III involved 110 participants who had participated in Phases I or II and re-consented for follow-up. A total of 204 participants provided mobility data. Individuals were eligible to participate if: (1) they were between 20 and 30 years old; (2) had participated in the 2019 annual AHRI HIV surveillance round; (3) resided in the southern AHRI HIV surveillance area; (4) were willing to participate in the study; and (5) owned a compatible smartphone with a minimum RAM  of 1GB, sufficient free space for the installation of the study app (PHASES II and III). Eligibility was not limited by HIV status; the study aimed to include both people living with HIV and those who were not.

\begin{figure}[htbp]
\centering
\includegraphics[width=1\textwidth]{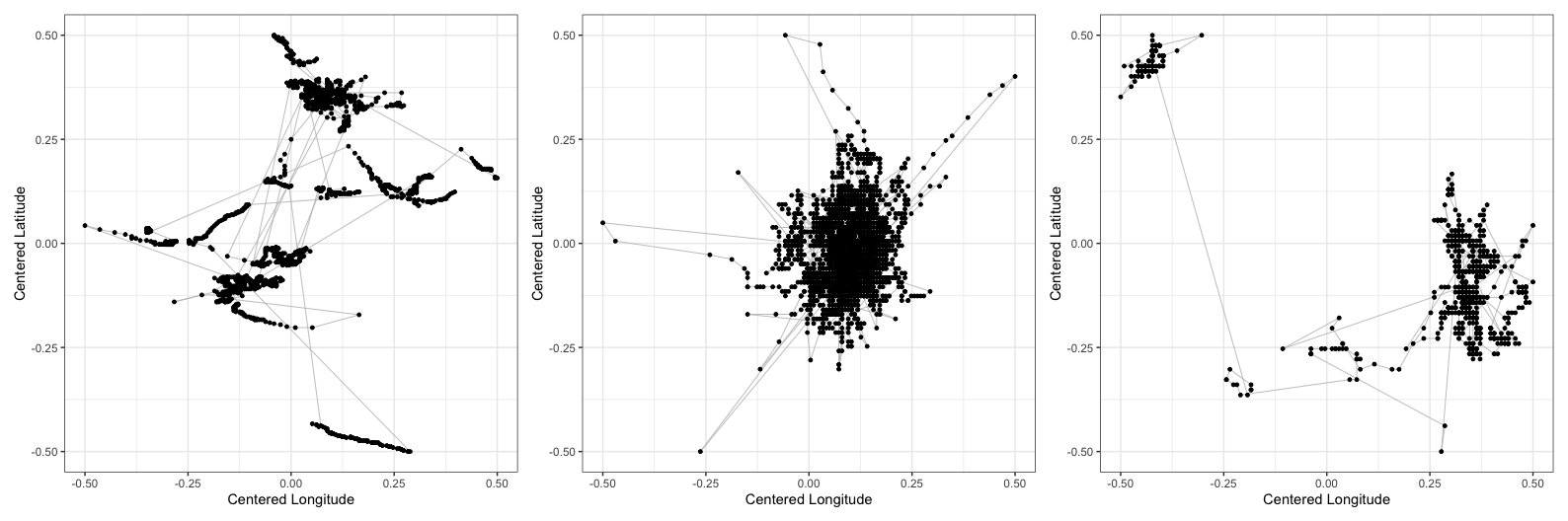}
\caption{Example spatial trajectories recorded for three individual devices over a two-day period in the Sesikhona GPS study. Latitude and longitude coordinates are shifted and rescaled to maintain privacy of the study participants.} 
 \label{fig:traj}
\end{figure}

Figure \ref{fig:traj} illustrates examples of movement between locations visited by three study participants over a two-day period. We observe that the mobility patterns of these study participants alternate between shorter and longer trips, departing and returning to several key areas after visiting other locations. We note that the spatial distribution, extent, and duration of the trips vary considerably among study participants.

\begin{figure}[htbp]
\centering
\includegraphics[width=1\textwidth]{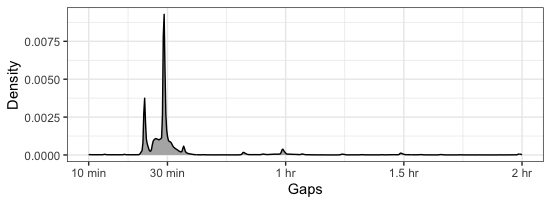}
\caption{Distribution of time intervals between consecutive GPS locations in the Sesikhona GPS study.} 
 \label{fig:gaps}
\end{figure}

The GPS locations of each study participant were recorded in variable time intervals of differing lengths. This was not due to the design of the Sesikhona study. Instead, several practical reasons explain why time intervals of irregular lengths occur. These include technological limitations (e.g., GPS devices running out of power), heterogeneous built environments that hinder GPS devices from obtaining a location (e.g., skyscrapers in downtown areas or buildings without windows and WIFI coverage), and human behavioral factors (e.g., study participants disabling their GPS devices near certain sensitive locations). Shorter time intervals between consecutive GPS locations generally allow for reasonable inferences about unrecorded movement by a study participant between the two recorded locations. However, longer time intervals complicate the reliable determination of a study participant's location. We define a gap as the duration during which the interval between two consecutive GPS locations is too long to make an informed determination about the unobserved locations visited by the study participant. In Figure \ref{fig:gaps}, we present the distribution of time intervals, which range from 10 minutes to 2 hours, in the Sesikhona data. The distribution peaks at approximately 30 minutes, with additional local modes observed near 25 minutes and at 1 hour. For this reason, we consider the time intervals between consecutive GPS locations to be gaps if their duration exceeds 30 minutes.

Table \ref{tab:gpssummary} presents summaries of mobility measures categorized by age group and sex for the mobility patterns of the participants in the Sesikhona GPS study. These measures include: (a) time spent inside, defined as the total duration, expressed in days, when a study participant was located within the AHRI study area; (b) distance traveled inside, defined as the average distance traveled per day, expressed in kilometers, by a study participant within the AHRI study area; (c) time spent outside, defined as the total duration, expressed in days, when a study participant was located outside the AHRI study area; and (d) distance traveled outside, defined as the average distance traveled per day, expressed in kilometers, by a study participant outside the AHRI study area. We note that gaps in the spatiotemporal trajectories of study participants were excluded from the calculation of these measures. The age of each participant was determined based on their age at the time of the longest recorded observation, as the study spans multiple years. Table \ref{tab:gpssummary} shows that on average study participants older than 26 tend to be more mobile compared to younger study participants. Men and women seem to have comparable levels of mobility inside and outside the AHRI study area. When outside the study area, the study participants are mobile compared to the time periods when inside the study area.

{\scriptsize
\begin{table}[htbp]
\centering
{\begin{tabular}{llccccc} \toprule
 & Age &  & Time Spent& Distance Traveled & Time Spent & Distance Traveled\\
Sex & Group & N\textsuperscript{a} & Inside\textsuperscript{b} & Inside \textsuperscript{b} & Outside\textsuperscript{b} & Outside\textsuperscript{b}\\ \midrule
Women & 20-26 & 62 & 32.44 $\pm$ 43.55 & 14.44 $\pm$ 19.30 & 8.18 $\pm$ 17.11 & 35.42 $\pm$ 33.51\\
Women & 27-34 & 69 & 57.00 $\pm$ 73.66 & 14.89 $\pm$ 41.21 & 7.98 $\pm$ 15.57 & 74.59 $\pm$ 105.56\\\midrule
Men & 20-26 & 40 & 27.94 $\pm$ 37.92 & 14.08 $\pm$ 8.91 & 7.01 $\pm$ 13.53 & 72.67 $\pm$ 112.26\\
Men & 27-34 & 33 & 45.26 $\pm$ 71.00 & 15.66 $\pm$ 12.60 & 8.50 $\pm$ 18.31 & 68.29 $\pm$ 128.10\\\bottomrule
\end{tabular}}
\caption{Summary mobility measures for the Sesikhona GPS Study. \textsuperscript{a} Sample size.\textsuperscript{b}Average $\pm$ standard deviation. The measures refer to time spent and distance traveled inside and outside the AHRI study area.}
\label{tab:gpssummary}
\end{table}
}

\section{Methods} \label{sec:methods}

In this section we present our methodological framework: a new method for the determination of the HIV status for participants in a longitudinal surveillance cohort (Section \ref{sec:imputationhiv}), a method for estimating HIV prevalence in a grid that covers a target region (Section \ref{sec:localhiv}), a method for activity space estimation from GPS data (Section \ref{sec:activityspace}) and our procedure for estimation of contextual exposure to HIV (Section \ref{sec:contextualexposure}) based on activity spaces.

\subsection{Imputation of the HIV status} \label{sec:imputationhiv}

Our method of estimating contextual exposure to HIV is based on local estimates of HIV prevalence throughout the AHRI study area. These estimates are based on the knowledge of the HIV status of each study participant in the AHRI demographic surveillance cohort eligible for HIV testing. Although study participants are eligible for periodic tests, they are often unavailable for their scheduled surveillance rounds due to work commitments, illness, transportation costs, frequent migration, and fear of stigma or discrimination \citep{tanser-et-2015,dobra-et-2017}. Some HIV tests may yield invalid results. This results in HIV surveillance records that lack information about the HIV status of a large proportion of study participants for periods of 2, 3, 4, or more years \citep{larmarange-et-2015}. 

\begin{table}[htbp]
{\scriptsize
\centering
{\begin{tabular}{llrrrr} \toprule
&& \multicolumn{4}{l}{HIV tests} \\ \cmidrule{3-6}
Sex & Quantile & Only Negative\textsuperscript{a} & Only Positive\textsuperscript{b} & Negative and Positive\textsuperscript{c} & No Valid Test\textsuperscript{d}\\ \midrule
Men & 2.5\% & 10 & 591 & 343 & 158\\
    & 50\% & 1,490 & 3,943 & 1,589 & 2,302\\
    & 97.5\% & 5,529 & 8,242 & 4994 &8,580\\ \midrule
Women & 2.5\% & 5 & 479 & 334 & 138\\
      & 50\% & 640 & 3,499 & 1,170 & 2,292\\
      & 97.5\% & 5,196 & 8,119 & 4692 & 8,514 \\ \bottomrule
\end{tabular}}
\caption{Periods of time measured in days when the HIV status of study participants is unknown. \textsuperscript{a}Number of days elapsed between the date of the study participant's latest HIV negative test and the date when the study participant left the surveillance cohort. \textsuperscript{b}Number of days elapsed between the date when the study participant joined the surveillance cohort and their earliest HIV positive test. \textsuperscript{c}Number of days elapsed between the study participant's latest HIV negative test and their earliest HIV positive test. \textsuperscript{d}Number of days elapsed between the date when the study participant joined the surveillance cohort and the date when the study participant left the study cohort.}
\label{tab:exposureduration}
}
\end{table}

To date, HIV status imputation methods have focused on determining the date of HIV seroconversion for individuals who have undergone repeated HIV testing and have experienced seroconversion. The most popular imputation method considers the seroconversion date to be the midpoint between the last negative and first positive test dates of the participant. However, midpoint imputation has been shown to be less desirable because HIV seroconversion is unlikely to occur at the midpoint between negative and positive tests, due to the influence of several key HIV risk factors such as sex, age, and migration \citep{skar-et-2013,dobra-et-2017}. For this reason, \citet{vandormael-et-2018} recommends using random-point imputation between the participant's last negative and earliest positive test dates. In subsequent work, \citet{vandormael-et-2020} proposed the G-imputation approach, which generates HIV seroconversion dates based on individual-level time-dependent and time-independent covariates. However, this method is only applicable to repeat HIV testers who become HIV positive. These study participants represent only a portion of the AHRI HIV surveillance cohort. Other portions involve:  (1) study participants who never tested positive for HIV, even though they tested negative at least once; (2) study participants who recorded only HIV positive tests without testing negative; and (iii) study participants who never recorded a valid HIV test. The date of seroconversion for study participants who tested positive for HIV but never tested negative cannot be inferred using the mid-point, random-point, or any existing imputation methods because there is no known date when these participants were HIV negative. Study participants who have never tested positive for HIV might seroconvert during their exposure period after their last HIV negative test. 

The method we propose offers a consistent framework for inferring the HIV status of all study participants throughout the duration of their exposure period. Our method allows for the possibility that any study participant might seroconvert during exposure periods when their HIV status is unknown. However, it is constrained to ensure that the HIV seroconversion date is consistent with the dates of any valid HIV negative or any HIV positive tests. Additionally, our method takes into account the likely possibility that a study participant might never seroconvert. As shown in Table \ref{tab:exposureduration}, the exposure period during which study participants have an unknown HIV status varies considerably irrespective of their record of valid HIV tests. This provides crucial empirical evidence for the need to develop a procedure to determine HIV status.

Our method is based on the assumption that HIV prevalence and HIV incidence estimates are available for each time period of interest and for each relevant age group (e.g., $[5-20)$,$[20-24)$,$\ldots$,$[50-54)$) separately for men and women. The process of imputing the seroconversion date (if any) for a single study participant is as follows.
Let $A_t$ ($t=1,2,\ldots,T$) be a random variable that takes the value $1$ if the study participant was HIV positive during the time period $t$, and the value $0$ if the study participant was HIV negative during the same time period. The probability that a study participant is HIV positive during the time period $t$ is represented by the prevalence of HIV $\mu_t=\Pr(A_t=1)$. The incidence for the time period $t$ is the probability that a study participant becomes HIV positive during the time period $t$, given that they were HIV negative in the previous time period:
$$
 \lambda_t = \Pr(A_t = 1\mid A_{t-1}=0),
$$
\noindent for $t=2,\ldots,T$. The prevalence $\mu_t$ and the incidence $\lambda_t$ are values associated with the age group of the study participant during the period of time $t$, as well as with their sex.

We assume that the study participant is known to be HIV negative during the time period $t-1$ ($t\ge 2$), based either on an HIV test or on a previous imputation step. We want to impute their HIV status for the period of time $t$, assuming that it is currently unknown. If $t\le T-1$, we also assume that their HIV status is unknown during the time period $t+1$. To perform the imputation, we sample from a Bernoulli distribution with probability of success $\lambda_t$.

Next, we assume that the HIV status of the study participant is unknown during the time period $t-1$ and that their HIV status is known to be positive during the time period $t$ ($t\ge 2$), i.e., $A_t=1$. If $t\ge 3$, we also assume that the HIV status of the study participant is unknown during the period of time $t-2$. Using the Bayes' rule, we find that the conditional probability of the study participant being HIV negative during the time period $t-1$ is the following.
\begin{eqnarray}
\Pr(A_{t-1}=0\mid A_t=1) & = & \frac{1-\mu_{t-1}}{\mu_t}\lambda_t. \label{eq:neggivenpos}   
\end{eqnarray}
The HIV status of the study participant for the time period $t-1$ is imputed by sampling from a Bernoulli distribution with probability of success \eqref{eq:neggivenpos}.

Now we consider the case where the HIV status of the study participant is unknown during the time period $t-1$, but is known to be HIV negative during the time period $t-2$ ($A_{t-2}=0$), and is also known to be HIV positive during the time period $t$ ($A_t=1$), where $t\ge 3$. The imputation of the HIV status during the time period $t-1$ can be conducted by sampling from a Bernoulli distribution with probability of success $\Pr( A_{t-1}=1\mid A_{t-2}=0,A_t=1)$ equal to:
\begin{eqnarray*}
 \frac{\Pr(A_t=1 \mid A_{t-2}=0, A_{t-1}=1)\Pr(A_{t-1}=1\mid A_{t-2}=0)}{\Pr(A_t=1\mid A_{t-2}=0)}.
\end{eqnarray*}
We have $\Pr(A_t=1 \mid A_{t-2}=0, A_{t-1}=1) = 1$, $\Pr(A_{t-1}=1 \mid A_{t-2}=0) = \lambda_{t-1}$ and $\Pr(A_t=1\mid A_{t-2}=0) = \lambda_{t-1} + \lambda_t (1-\lambda_{t-1})$.
We obtain:
\begin{eqnarray}
\Pr( A_{t-1}=1\mid A_{t-2}=0,A_t=1) & = & \frac{1}{1+\lambda_t\left(\frac{1-\lambda_{t-1}}{\lambda_{t-1}}\right)}. \label{eq:posgivennegpos}
\end{eqnarray}

The imputation of the HIV status of a study participant who tested negative but never tested positive is performed sequentially, period by period, moving forward in time, starting with the period of time in which they tested negative. The HIV status of the study participant for the next period of time $t$ is sampled from a Bernoulli distribution with probability of success $\lambda_t$. We stop sequential imputation if the HIV status of the study participant is sampled as positive. In this case, the study participant remains HIV positive for the remaining time periods until the end of their exposure period.

The imputation of the HIV status of a study participant that tested positive but never tested negative is performed sequentially time period by time period by moving backward in time starting with the time period when they tested positive. The HIV status of the study participant for the time period $t-1$ given that their HIV status is positive in the time period $t$ is sampled from a Bernoulli distribution with probability of success \eqref{eq:neggivenpos}. We stop sequential imputation if the HIV status of the study participant is sampled as negative. When this happens, the study participant remains HIV negative for the remaining time periods until the beginning of their exposure period.

The imputation of HIV status for a study participant that tested negative in an earlier time period and tested positive in a later time period proceeds by imputing their HIV status sequentially time period by time period and moving forward in time from the time period of their last HIV negative test or by moving backward in time from the time period of their first HIV positive test. If moving forward in time, imputation for the time period $t$ given that HIV status is unknown in the time period $t+1$ is performed by sampling from a Bernoulli distribution with probability of success $\lambda_t$. If the HIV status of the study participant is sampled as positive, the HIV status is set to positive for the rest of the time periods that follow the time period $t$. If we move back in time, the imputation for the time period $t-1$ since HIV status is unknown for the time period $t-2$ is carried out by sampling a Bernoulli distribution with probability of success \eqref{eq:neggivenpos}. If the HIV status is sampled as negative, the HIV status is set to negative for the rest of the time periods that precede the time period $t-1$. Forward or backward imputation stops when we reach a time period for which the HIV status is still unknown, but the HIV status in the previous time period and the HIV status in the next time period are known either because they were previously imputed or because the study participant recorded a valid HIV test. Their HIV status in this period of time is imputed by sampling a Bernoulli distribution with probability of success \eqref{eq:posgivennegpos}.

Lastly, we consider imputing the HIV status of a study participant who never had a valid HIV test. In their first exposure period, the HIV status is imputed by sampling from a Bernoulli distribution with the probability of success equal to the estimated HIV prevalence for the study participant's age group and sex during that time period. In subsequent time periods, if the HIV status of the study participant was not imputed as HIV positive, the imputation of their HIV status is performed by sampling from a Bernoulli distribution with probability of success equal to the estimated HIV incidence for their age group and sex during that time period.

\subsection{Determining local HIV prevalence estimates} \label{sec:localhiv}

In this section, we show how to estimate the prevalence of HIV in a grid that encompasses the AHRI study area, dividing it into 44,937 grid cells, each measuring 100 by 100 meters. We start by imputing the HIV status of people who are part of the AHRI HIV surveillance cohort between 2011 and 2023 for the entire duration of their existence in the cohort using the methodology described in Section \ref{sec:imputationhiv}. For each cohort member and each calendar year, we determine the location of their homestead of residence from the AHRI demographic surveillance data. Individuals who reside outside the AHRI study area are excluded from the HIV prevalence calculations for the corresponding calendar years. 

For each grid cell $i$ and the residential homestead $j$, we calculate the distance $d_{i,j}$ between the centroid of the grid cell and the location of the homestead. Distances $d_{i,j}$ are transformed into spatial weights $w_{i,j}$ using a two-dimensional Gaussian kernel with a search radius of 3-km \citep{waller-gotway-2004}:
\begin{equation}
w_{i,j} = \exp\left(- \frac{d_{i,j}^2}{2 s^2}\right),
    \label{eq:spatialweights}
\end{equation}
\noindent where $s \approx 1.165$. This value of the standard deviation $s$ indicates that the probability of distances from the centroid of a grid cell to a homestead exceeding 3 km is 0.01. 

The prevalence of HIV at the level of each grid cell $i$ in a calendar year is calculated as the ratio of the sum of the weights \eqref{eq:spatialweights} between the grid cell $i$ and the homesteads $j$ in which the people who are HIV positive in that calendar year reside, and the sum of the weights between the grid cell $i$ and the homesteads $j$ in which all individuals who are part of the AHRI HIV surveillance cohort reside \citep{friis-sellers-2009}. This method is appropriate for the scattered distribution of the population that resides in the AHRI study area because it does not impose static geographical boundaries on the relevant spatial locations. Instead, it uses the location of each homestead to derive a local estimate of HIV prevalence that captures local variations and is robust to the effects of noise.

\begin{figure}[htbp]
\centering
\includegraphics[width=1\textwidth]{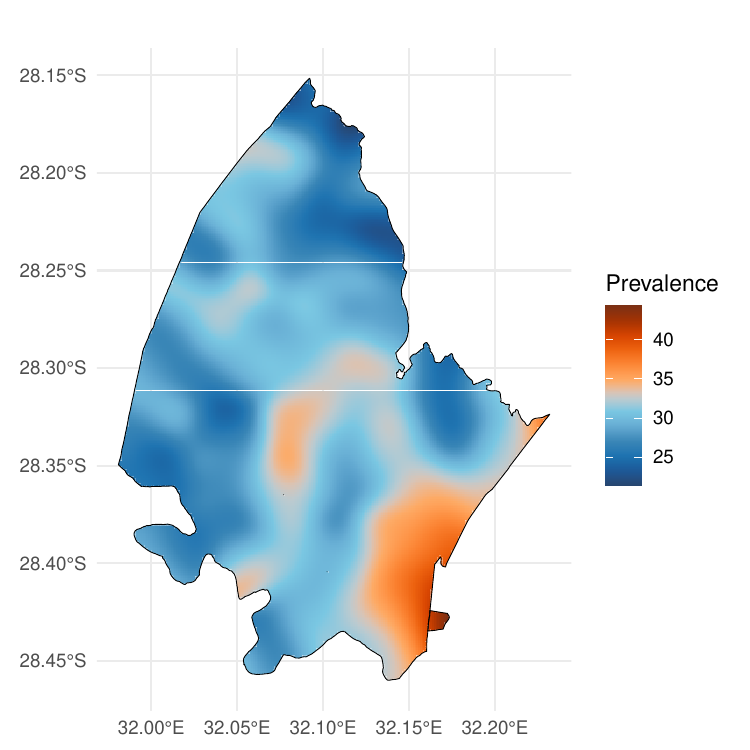}
\caption{Local estimates of HIV prevalence in the AHRI study area. Prevalence is expressed as a percentage.} 
 \label{fig:hivprevalence}
\end{figure}

A map illustrating the grid cell level estimates of HIV prevalence is presented in Figure \ref{fig:hivprevalence}. The highest prevalence of HIV, exceeding 40\%, is found in the town of Mtubatuba and its surrounding neighborhoods. This is expected, as this location has the highest population density, where most commercial activities occur in the rural region of the AHRI study area. The map also shows a significant spatial variation in HIV prevalence within the AHRI study area, which is consistent with previous findings \citep{tanser-et-2009}.

\subsection{Activity space estimation from GPS data} \label{sec:activityspace}

We present our method for estimating the activity spaces of the participants in the Sesikhona GPS study. We assume that study participants spend most of their time within a reference time frame $[t_{\min},t_{\max}]$ in a spatial observation window $\mathcal{W}\subset \mathbb{R}_{+}^2$, which is divided into a set of grid cells $\mathcal{G} = \{ G_1,\ldots,G_N\}$. The spatiotemporal trajectory of an example study participant is represented as a curve
\begin{eqnarray} \label{eq:trajectory}
 X^{[t_{\min},t_{\max}]} & = & \{ X(t) = (x_1(t), x_2(t)): t\in [t_{\min},t_{\max}]\} \subseteq \mathcal{W},
\end{eqnarray}
\noindent where $x_1(\cdot)$ and $x_2(\cdot)$ represent the longitude and latitude coordinates, respectively, and $X(t)$ is the location visited by this individual at time $t$. Each location $X(t)$ on the curve $X^{[t_{\min},t_{\max}]}$ belongs to a grid cell $G(t)\in \mathcal{G}$.

We represent the GPS data for a study participant as spatial locations on their spatiotemporal trajectory \eqref{eq:trajectory} recorded at timestamps $t_{\min}=t_1<t_2<\ldots<t_n=t_{\max}$ which are realizations of a random variable $T$ on $[t_{\min},t_{\max}]$:
\[
(X_{i},\, t_{i}) \ \in\  \mathcal{W}\times [t_{\min},t_{\max}],\qquad
i = 1,\dots,n,
\]
where \(X_{i}=X(t_{i})\) is the spatial coordinate of the \(i\)-th GPS observation. We denote by $\mathcal{T}$ the portion of the reference time $[t_{\min},t_{\max}]$ that does not belong to any gap in the GPS records of the example study participant. We note that we do not remove any GPS observations in the process of removing the gaps that exist in the reference time frame $[t_{\min},t_{\max}]$. Instead, we remove time intervals \([t_i, t_{i+1}]\) defined by consecutive time stamps with a difference \(t_{i+1} - t_i\) greater than a predefined threshold of 30 minutes, which was empirically justified in Section \ref{sec:sesikhonastudy}. After removing all such gaps, the remaining time domain consists of disjoint, contiguous segments:
\[
\mathcal{T} = \bigcup_{k=1}^{K} \bigcup_{i \in \mathcal{I}_k} [t_i, t_{i+1}],
\]
where each \(\mathcal{I}_k \subset \{1,2,\ldots,n-1\}\) represents a consecutive block of indices found between two gaps, and \(K\) denotes the total number of non-gap segments. Although it might seem reasonable to consider that the distribution of $T$ is uniform in $[t_{\min},t_{\max}]$ to have the same chance of recording any visited location during the reference time frame, the existence of gaps only allows us to assume that the distribution of $T$ is uniform in each interval $[t_i,t_{i+1}]$, $i\in \mathcal{I}_k$ for any $k\in\{1,2,\ldots, K\}$.

The activity distribution of a study participant across the grid cells $\mathcal{G}$ is $\pi = (\pi_1,\ldots,\pi_N)$, where $\pi_j$ denotes the proportion of time that this study participant spent in cell $G_j\in \mathcal{G}$ during $\mathcal{T}$:
\begin{equation} \label{eq:activityunif}
 \pi_j = \Pr(G(T)=G_j), \quad \mbox{for } j=1,\ldots,N.
\end{equation}
\citet{dong2020statistical} introduce and study the asymptotic properties of the following estimator of the activity distribution $\pi$, which they call the conservative proportional-time estimator (CPT, henceforth):
\begin{equation}
\widehat{\pi}_{j} = \frac{\sum_{k=1}^{K}\sum_{i\in \mathcal{I}_k}(t_{i+1}-t_{i})\mathbf{1} (g_{i} = g_{i+1} = G_k)}{\sum_{j=1}^{K+1}\sum_{i\in \mathcal{I}_k} \mathbf{1} (g_i = g_{i+1})},\quad \mbox{for } j=1,\ldots,N.
\label{eq:conservativeactivity}
\end{equation}
The estimator \eqref{eq:conservativeactivity} considers only time intervals linked to consecutively observed GPS locations that do not contain gaps and during which the study participant remained within a specific cell of the grid. Time intervals between consecutively observed GPS locations that are gaps or in which the study participant transitions from a grid cell to another are discarded.

We extend the estimator \eqref{eq:conservativeactivity} from a single study participant to a group of participants denoted by $\{1,\ldots, k\}$. For each participant $i$ in this group, let $\hat{\pi}_i = (\hat{\pi}_{1,i}, \ldots, \hat{\pi}_{N,i})$ denote the estimated individual-level activity distribution obtained by the CPT estimator \eqref{eq:conservativeactivity}.
For any subgroup of participants indexed by a set $C \subseteq\{1,\ldots, k\}$, the subgroup-specific activity distribution is given by
\begin{equation}
\hat{\pi}_{C} = \frac{\sum_{i=1}^k \hat{\pi}_i \mathbf{1}\{i\in C\}}{\sum_{j=1}^N \sum_{i=1}^k \hat{\pi}_{j,i}\mathbf{1}\{i\in C\}}.
\label{eq:conservativeactivity_sub}
\end{equation}
The estimator \eqref{eq:conservativeactivity_sub} incorporates the total time contributed by the participants in $C$ for each cell of the grid. Normalization in the denominator ensures that $\hat{\pi}_{C}$ is a valid probability distribution on the cells of the grid. 
%Then, the collective activity distribution across all study participants is defined as \begin{equation} \hat{\pi}_{\text{all}} = \frac{\sum_{i=1}^k \hat{\pi}_i}{\sum_{j=1}^N \sum_{i=1}^k \hat{\pi}_{i,j}}. \label{eq:conservativeactivity_all} \end{equation}

Given a study participant’s activity distribution over grid cells, we define the activity space of that person $\mathcal{AS}$ as the set of grid cells in which they spent time: 
\begin{equation}
    \mathcal{AS} =  \bigcup_{\{j:\pi_j>0\}}\{ G_j\}.
    \label{eq:activityspace}
\end{equation}
For any subgroup of participants indexed by a set $S \subseteq\{1,\ldots, k\}$, the collective activity space is defined as the union of the individual activity spaces:
\begin{equation}
    \mathcal{AS}^S = \bigcup_{i\in S} \mathcal{AS}^i,
    \label{eq:activityspace_sub}
\end{equation}
where $\mathcal{AS}^i$ is the activity space of an individual $i$.

For any \(\gamma \in (0,100]\), the level-\(\gamma\) activity space is defined as the subset of grid cells with the smallest number of elements in which the study participant spends at least \(\gamma\%\) of their total time, denoted by \(\mathcal{AS}_{\gamma}\). Specifically, define
\begin{equation}
\mathcal{Q}_{\gamma} = \left\{ Q \subseteq \mathcal{G} : \sum_{\{j : G_j \in Q\}} \pi_j \ge \gamma\% \right\},
\label{eq:gammatarget}
\end{equation}
where \(\pi_j\) represents the proportion of time spent in grid cell \(G_j\).
Then \(\mathcal{AS}_{\gamma}\) is the element of \(\mathcal{Q}_{\gamma}\) that satisfies
\[
|\mathcal{AS}_{\gamma}| \le |Q| \quad \text{for all } Q \in \mathcal{Q}_{\gamma}.
\]
If multiple subsets in \(\mathcal{Q}_{\gamma}\) achieve the same minimal cardinality, we select the one whose total time-spent proportion, \(\sum_{\{j : G_j \in Q\}} \pi_j\), is maximal.

For any $\gamma\in (0,100]$, the estimation of $\mathcal{AS}_{\gamma}$ from the GPS data is based on the CPT estimator in \eqref{eq:conservativeactivity}. In order to estimate $\mathcal{AS}_{\gamma}$, we order the grid cells in decreasing order of their estimated time spent proportion $\{\widehat{\pi}_j\}_{j=1}^N$. Starting with the first cell in this ordering, we add cells to $\mathcal{AS}_{\gamma}$ until the cumulative time spent proportion in \eqref{eq:gammatarget} reaches the desired level $\gamma$. 

To capture the magnitude and structural complexity of human movement in a comparable form,  we use the number of distinct cells in the grid visited at different levels of the activity space ($|\mathcal{AS_{\gamma}}|$) as a measure of mobility. Traditional metrics such as total distance traveled capture how much individuals move, but do not reflect where they move and how heterogeneous their mobility patterns are. In contrast, $|\mathcal{AS_{\gamma}}|$ provides a spatially explicit measure of movement diversity and dispersion. It is less sensitive to random GPS noise or repetitive small-scale movements under high recording-frequency  (such as walking within the same compound) that can artificially inflate total distance, yielding a more stable and behaviorally meaningful indicator of mobility.

To relate $\mathcal{AS}$ to specific geographic contexts, we distinguish between regions inside and outside the AHRI study area. Let $\mathcal{U}_{\text{in}}$ denote the AHRI study area and $\mathcal{U}_{\text{out}} = \mathcal{U}_{\text{in}}^c$ its complement within South Africa.
Then $\mathcal{AS}, \mathcal{AS}(\mathcal{U}_{\text{in}}), \mathcal{AS}(\mathcal{U}_{\text{out}})$ respectively represent the activity space in South Africa, within the AHRI study area, and outside the study area but within South Africa. 
Analogously we define $\mathcal{AS}_{\gamma}(\mathcal{U}_{\text{in}})$ and $\mathcal{AS}_{\gamma}(\mathcal{U}_{\text{out}})$ to represent activity spaces inside and outside the AHRI study area given a level $\gamma$.

\begin{figure}[htbp]
\centering
\includegraphics[width=1\textwidth]{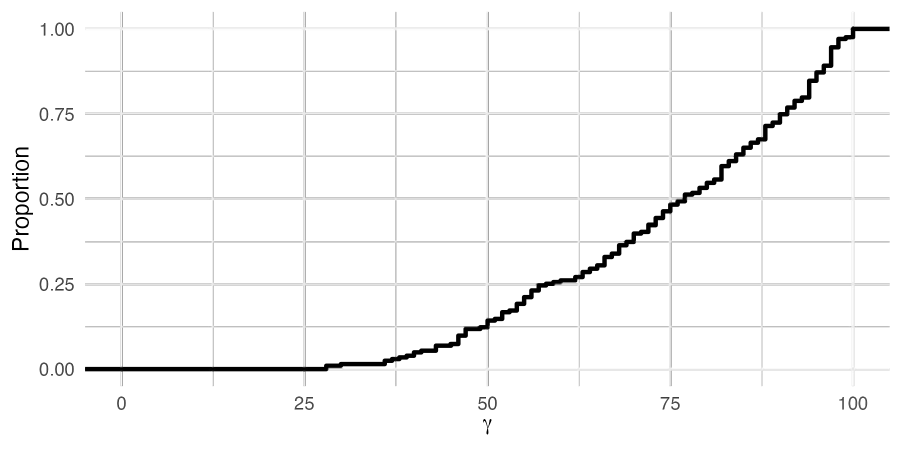}
\caption{The empirical cumulative distribution of the smallest \(\gamma\) at which \(|\mathcal{AS}_{\gamma}|>1\).}
\label{fig:num_grid_gt1_cdf}
\end{figure}

We also define home activity space as the smallest set of grid cells that accounts for at least 50\% of an individual’s total observed time ($\mathcal{AS}_{50}$). The 50\% level is a pragmatic choice that aligns with the daily time budget in which home typically dominates non-working hours, giving an interpretable core that is comparable across participants.
Figure~\ref{fig:num_grid_gt1_cdf} shows the proportion of participants with $|\mathcal{AS}_{\gamma}|>1$ for each level of activity space $\gamma$.
At the 50\% level, for most people, the activity spaces contain only one cell, and therefore our choice of home activity space level is quite conservative. Nevertheless, this provides a natural reference against which broader mobility patterns can be compared and serves as a principled substitute for a single static home location.

\subsection{Estimation of the contextual exposure to HIV} \label{sec:contextualexposure}

The contextual exposure to HIV for the participants in the Sesikhona study is determined on the basis of their GPS spatio-temporal trajectories as follows. The GPS records of the study participants are grouped in locations within or outside of the AHRI study area. The locations within the AHRI study area are used to determine the activity distributions and activity spaces of the study participants using the methods described in Section \ref{sec:activityspace}. The locations outside the AHRI study area are mapped in the 52 districts of South Africa, and the corresponding activity distributions and activity spaces are determined at the country level.  

Contextual exposure to HIV when a study participant is within the AHRI study area is calculated as a weighted average of the grid cell-level estimates of HIV prevalence introduced in Section \ref{sec:localhiv} with weights representing the proportion of time spent by the study participant in each grid cell. Contextual exposure to HIV when a study participant is outside the AHRI study area is determined as a weighted average of district-level estimates of HIV prevalence obtained from \citet{ihme-2019,ihme-2021} with weights representing the proportion of time spent by a study participant in each district. 

In the sequel we focus on four individual-level measures of contextual exposure to HIV: general exposure ($\mathcal{E}_{\text{overall}}$), exposure within the AHRI study area ($\mathcal{E}_{\text{in}}$), exposure outside the AHRI study area ($\mathcal{E}_{\text{out}}$) and exposure within the participant’s home activity space $\mathcal{AS}_{50}$ ($\mathcal{E}_{\text{home}}$).
$\mathcal{E}_{\text{overall}}$ is determined as a weighted average of the contextual exposure to HIV when a study participant is within and outside the AHRI study area, and therefore might rely on the grid cell-cell and district level estimates of HIV. 
Similarly, $\mathcal{E}_{\text{home}}$ is defined as a weighted average of contextual exposure over $\mathcal{AS}_{50}$.

\section{Analysis of the Sesikhona GPS data} \label{sec:analysis}

Our analysis focuses on determining the relationship between the  mobility of the participants in the Sesikhona GPS study and their demographic characteristics, such as sex and age \--- see Section \ref{sec:mobilitypatterns}. In Section \ref{sec:hivexposure}, we examine the differences between contextual exposure to HIV at home and the complete activity spaces of the study participants. We discuss the application of GPS-based mobility measures and HIV contextual exposure to identify individuals at high risk of HIV acquisition.

\subsection{Demographic variation in mobility patterns} \label{sec:mobilitypatterns}

Understanding the demographic differences in mobility behavior provides critical context for interpreting spatial exposure patterns. Because movement determines where individuals spend their time and which environments they encounter, differences in the structure of the activity-space between men and women — or between age groups— may translate into varying levels of HIV exposure. We compare the overall spatial distribution of men and women, along with their activity spaces, and then examine how demographic factors influence key mobility indicators.

We begin with a descriptive comparison of mobility, specifically the activity-space coverage, between male and female study participants. We compute the activity distribution 
$\hat{\pi}_i = (\hat{\pi}_{1,i}, \ldots, \hat{\pi}_{N,i})$ for each individual $i$ according to \eqref{eq:conservativeactivity}, where $\hat{\pi}_{j,i}$ denotes the estimated proportion of time that individual $i$ spent in grid cell $G_j$ after temporal gaps have been removed. Then, we find the activity distribution for women ($\hat{\pi}_{\mathcal{F}}$), men ($\hat{\pi}_{\mathcal{M}}$), and all study participants ($\hat{\pi}_{\mathcal{F}\cup\mathcal{M}}$) by \eqref{eq:conservativeactivity_sub}, where $\mathcal{F}$ and $\mathcal{M}$ denote the sets of indices corresponding to women and men, respectively. 

\begin{figure}[htbp]
\centering
\includegraphics[width=1\textwidth]{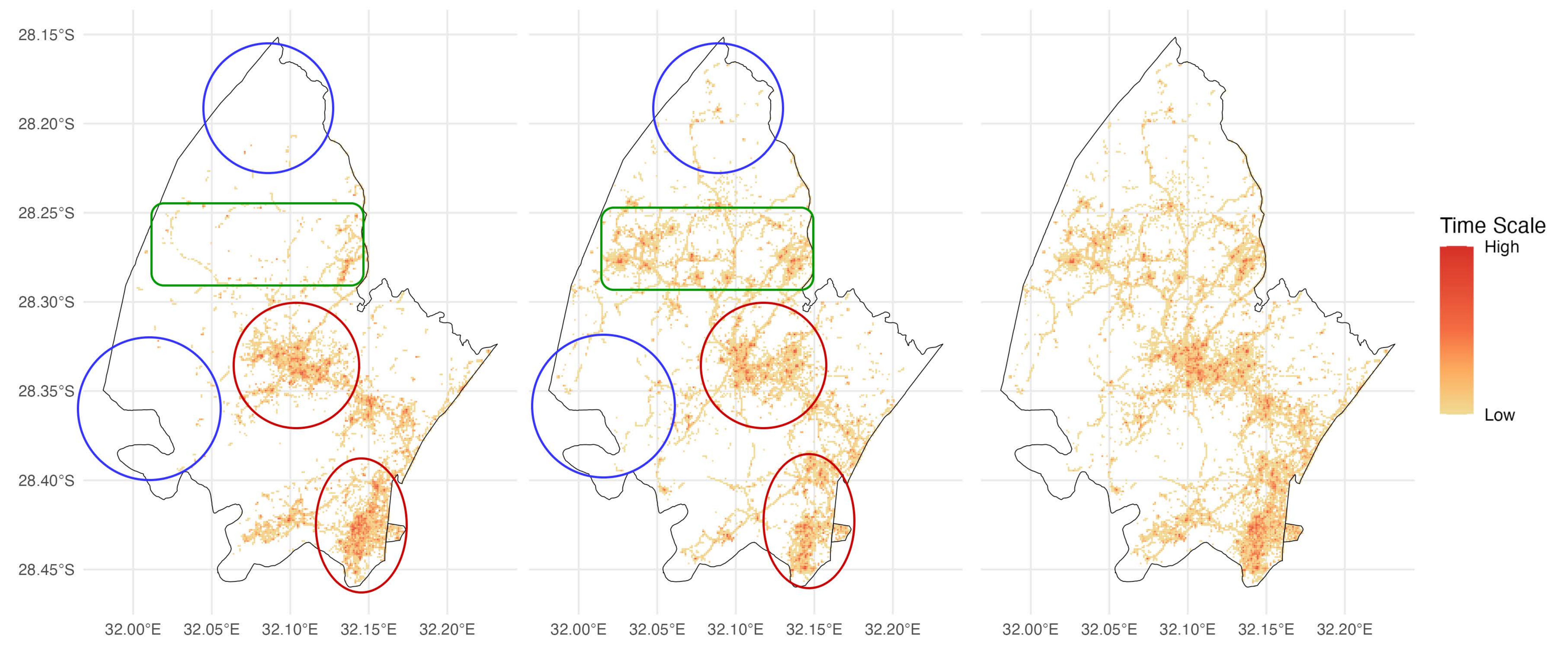}
\caption{Activity distribution within the AHRI study area for men $\hat{\pi}_\mathcal{M}$ (left), women $\hat{\pi}_\mathcal{F}$ (middle), and all participants $\hat{\pi}_{\mathcal{F}\cup \mathcal{M}}$ (right). White grid cells indicate no recorded time spent.} 
 \label{fig:time_inside}
\end{figure}

To visualize the activity distributions for each group of participants, we plot $\log(\hat{\pi}_S+\varepsilon)$ on spatial grids of the AHRI study area (Figure \ref{fig:time_inside}). Here, where $S\in \{\mathcal{F}, \mathcal{M}, \mathcal{F}\cup\mathcal{M}\}$ and $\varepsilon = 10^{-15}$ serve as a small offset to suppress extreme peaks, thereby improving the visibility of areas with low-activity and preventing the logarithm of zero from being taken.
Figure~\ref{fig:time_inside} indicates that both men and women spend the most time in several hotspots marked with red circles. Areas marked with blue circles indicate regions with little or no recorded activity, which correspond to sparsely populated or uninhabited areas. There are some clear differences in the mobility patterns of men and women. Men exhibit a more clustered movement pattern, with time concentrated in specific cells, whereas women demonstrate a more dispersed movement pattern, with significant activity levels spread across a broader set of grid cells, including those areas enclosed by the green rectangles. %Both panels indicate that there is no significant difference in the overlap of mobility patterns between women and men.
Although the activity distributions are normalized, the apparent larger spatial spread in women's activity distribution may be inflated by the fact that a greater number of women (N=91) than men (N=73) participated in the GPS study. To allow for a fair comparison, we randomly draw, without replacement, multiple subsets of mobility patterns associated with female study participants, where the size of each subset equals the number of male study participants—see the top panel of Figure \ref{fig:num_grid}.
Maps depicting activity distributions outside the AHRI study area can be found in Appendix \ref{app:time_outside}. 

%-------------#grid------------------
To quantify the spatial spread of the activity distribution among study participants, we determine the number of distinct grid cells visited at each activity space level \(|\mathcal{AS}_{\gamma}|\). Separately for women and men, i.e., for \(S\subseteq \{\mathcal{F},\mathcal{M}\}\), we calculate two measures: (i) \(|\mathcal{AS}^S_{\gamma}|\), the number of grid cells in the collective activity space; and (ii) \(\frac{1}{n_S}\sum_{i\in S}|\mathcal{AS}_{\gamma}^i|\), the average number of grid cells in the individual activity spaces. The first measure focuses on the number of distinct grid cells that the group as a whole needs to cover \(100\gamma\%\) of the total time spent by the group and accounts for overlap between individual activity spaces. The second measure represents the average number of cells that one person in a group needs to cover \(100\gamma\%\) of their time. This ignores the overlap within the group but reflects the footprint per-person.
The top panel of Figure~\ref{fig:num_grid} displays the first measure at $\gamma\in [0,100]$, while the bottom panel of Figure~\ref{fig:num_grid} presents the second measure at $\gamma\in [0,100]$.
To account for the imbalanced sex groups, the curve for women in the plot in the top panel is generated by randomly sampling women's activity distributions without replacement, ensuring that the sample size matches that of the men. Sampling was repeated 100 times, and the resulting curves were averaged. 
The curves increase rapidly with \(\gamma\) and are convex, indicating that capturing additional time shares becomes progressively more challenging with a small number of cells.
At the same level \(\gamma\), men require more grid cells than women in both summaries. This implies that men exhibit greater mobility compared to women.

\begin{figure}[htbp]
\centering
\includegraphics[width=\textwidth]{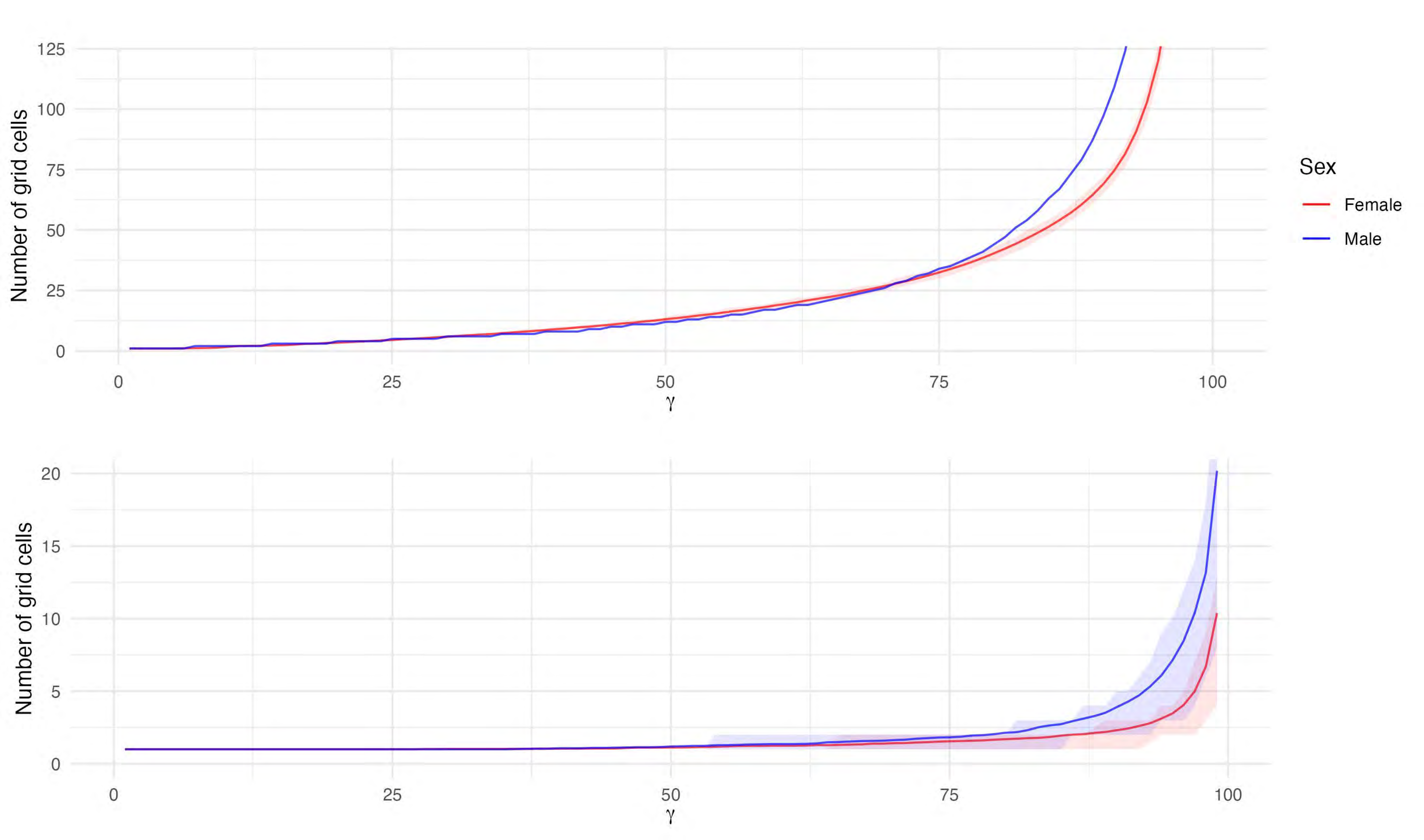}
\caption{Distribution of the number of grid cells across activity space levels for female and male participants within the AHRI study area: collective activity space (top panel) and average across the individual activity spaces (bottom panel). The shaded area represents the inter-quartile range (for the red curve in the top panel, this area is very narrow and may be hard to see).} 
 \label{fig:num_grid}
\end{figure}

\begin{figure}[htbp]
\includegraphics[width=1\textwidth]{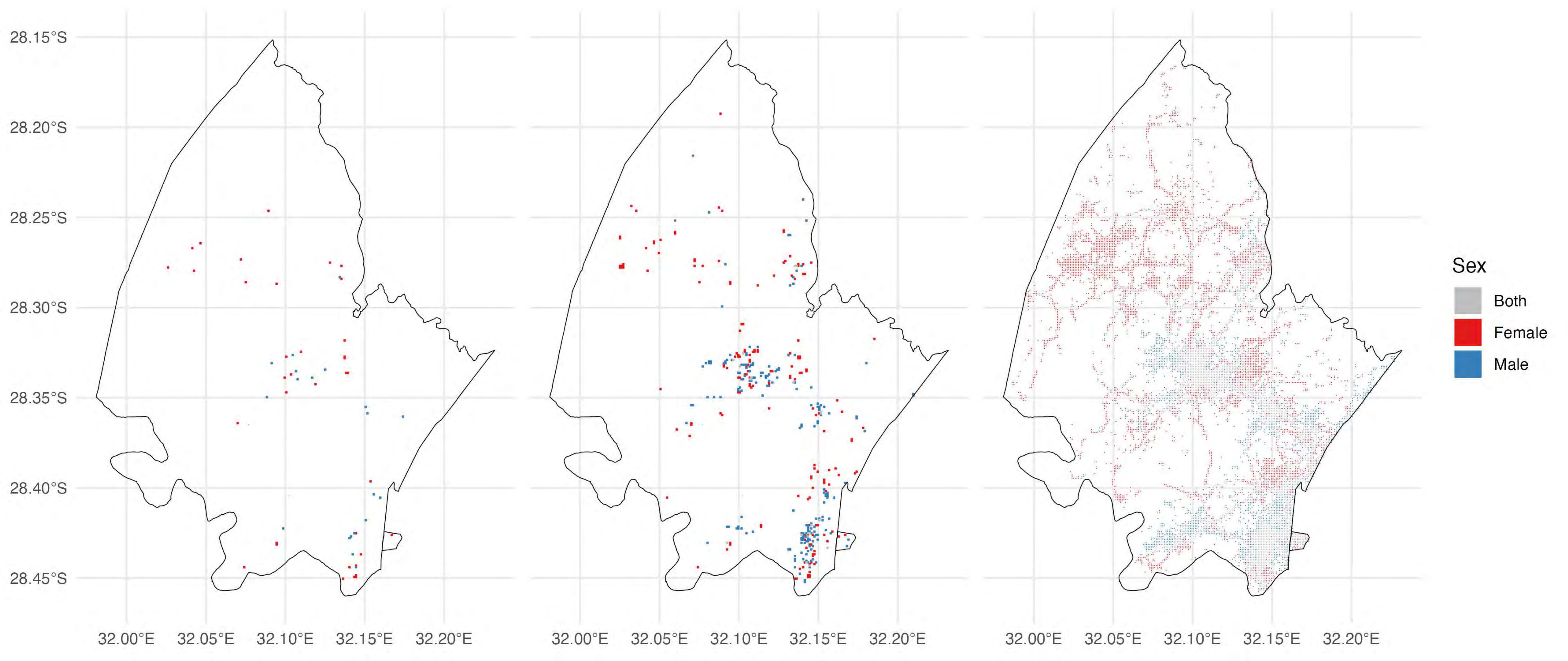}
\caption{Collective activity spaces of all individuals within the AHRI study area colored by sexes at different levels: $\mathcal{AS}^{\mathcal{F}\cup\mathcal{M}}_{65}(\mathcal{U}_{\text{in}})$ (left panel), $\mathcal{AS}^{\mathcal{F}\cup\mathcal{M}}_{95}(\mathcal{U}_{\text{in}})$ (middle panel), and $\mathcal{AS}^{\mathcal{F}\cup\mathcal{M}}_{100}(\mathcal{U}_{\text{in}})$ (right panel). The figure illustrates how grids visited by both men and women, as well as the collective activity space, expand with increasing levels.}
\label{fig:map_overlap}
\end{figure}

%-----------overlap------------------
% Since we found no apparent distinction in overlap within each sex, 
To better understand the common structure of male and female mobility, we investigated the spatial overlap between the activity spaces of women and men.
Figure~\ref{fig:map_overlap} illustrates the collective activity spaces of all participants at three levels of $\gamma$ (65\%, 95\% and 100\%), with grid cells colored according to sex. These levels were selected to capture various aspects of mobility. The 65\% level corresponds to the highest activity space threshold at which male and female grid cells do not overlap. The 95\% level represents the set of locations that encompasses nearly all regular and recurrent movements while excluding infrequent or potentially noisy points, such as long-distance trips or GPS errors. The 100\% level represents the entire activity space, encompassing all recorded locations, irrespective of their frequency. As $\gamma$ increases, the collective activity space expands, leading to a more pronounced spatial overlap between the movements of men and women. At the 100\% level, most of the cells in the grid visited by men are also visited by women; however, men do not use a noticeable subset of the cells visited by women. This pattern is due in part to the larger female cohort compared to the male cohort, but also highlights sex-specific differences in mobility, with women engaging in activities across a greater variety of locations.

%\subsubsection*{Quantitative assessment of demographic influences on mobility}
Up to this point, our analyses suggest that sex may influence differences in the mobility patterns of the study participants. We assess these differences and examine whether age influences mobility by modeling the number of grid cells within each individual’s activity space using a Poisson mixed-effects regression. For study participant $i$ at the activity space level $\gamma\in[50, 95]$, we fitted a log-linear model that includes a subject-specific random intercept:
\begin{eqnarray}
Y_{i\gamma} & \mid & b_i \sim \mathrm{Poisson}(\mu_{i\gamma},\phi),\nonumber \\
\log \mu_{i\gamma} & = & \beta_0 + b_i + \beta_1,\texttt{Male}_i + \beta{2}\gamma_i + \beta_3\texttt{Age}_i + \beta_4\texttt{Age}_i^2 + \beta_5\texttt{Age}_i^3, \label{eq:regression}\\
b_i &\sim& \mathcal{N}(0,\sigma_b^2),\nonumber
\end{eqnarray}
where \(Y_{i\gamma}\) represents the number of grid cells in $|\mathcal{AS}_\gamma^i|$, and \(\texttt{Male}_i\) takes the value of 1 for men and 0 for women. The random intercept \(b_i\) accounts for the repeated measurements of each individual across various levels $\gamma$. We include polynomial terms of degree 3 for age to capture its nonlinear effects on the outcome.
Parameter estimates and likelihood-ratio tests comparing the full model to reduced models—each excluding one covariate sequentially, are reported in Table \ref{tab:regression}.

\begin{table}[htbp]
\centering
{\begin{tabular}{lccc}
\hline
\textbf{Predictor} & \textbf{Estimate} & \textbf{Std. Error} & \textbf{LRT p-value\textsuperscript{a}} \\
\hline
 \texttt{Male} & 0.239 & 0.062 & $< 0.001$\\
 \texttt{Activity space level} ($\gamma$) & 0.032 & 0.001 & $< 0.001$\\
 \texttt{Age} & 0.008 & 0.062 & \multirow{3}{*}{0.980}\\
 \texttt{Age}$^{2}$ & 0.012 & 0.028 & \\
 \texttt{Age}$^{3}$ & -0.003 & 0.025 & \\
\hline
\multicolumn{4}{l}{\textit{Random effects (variance components):} $\sigma^2 = 0.162$} \\
\hline
\end{tabular}}
\label{tab:regression}
\caption{Poisson mixed-effects regression models predicting the number of grid cells $|\mathcal{AS}_\gamma^i|$ from sex, age, and activity space level, including random intercepts for individuals \--- see \eqref{eq:regression}. \textsuperscript{a}Each entry is the p-value from a likelihood-ratio test comparing the full model \eqref{eq:regression} to a reduced model that omits the covariate named in that row. For \texttt{Age}, the reduced model drops all polynomial terms in age jointly ($\texttt{Age}, \texttt{Age}^2, \texttt{Age}^3$). Small p-values indicate that removing the covariate worsens model fit.}
\end{table}

Higher levels of activity space were associated with a greater number of visited grid cells. The positive coefficient for \texttt{Male} indicates that, at the same activity space level, men tend to visit more locations than women, which aligns with the observations presented in Figure \ref{fig:num_grid}. We examined the contribution of age by testing the joint null hypothesis that all coefficients associated with age were equal to zero. Table \ref{tab:regression} shows that age does not significantly contribute to explaining the variation in the number of grid cells when gender and activity space are taken into account.

Together, the spatial and regression analyses indicate that the sex of the participants systematically influences the extent and structure of their’ mobility. These behavioral differences establish the foundation for understanding variations in contextual HIV exposure, which will be examined in the following section.

\subsection{Integrating HIV exposure with mobility analysis}\label{sec:hivexposure}

We aim to enhance our understanding of the relationship between movement behaviors and contextual HIV exposure by incorporating exposure estimation into our analyses. Traditionally, exposure has been measured at a study participant’s home location  based on the assumption that the residential environment adequately represents the individual's risk context \citep{entwisle-2007,kwan-2009}. However, this assumption may overlook exposures that occur outside the home environment, particularly for study participants who experience significant daily mobility. Based on GPS data, we define several exposure measures $\mathcal{E}_{\text{overall}}$, $\mathcal{E}_{\text{in}}$, $\mathcal{E}_{\text{out}}$, $\mathcal{E}_{\text{home}}$ to assess whether the inclusion of detailed mobility information impacts estimates of HIV exposure. 

In what follows, study participants must be distinguished based on whether their home activity spaces are located inside or outside the AHRI study area. This distinction accounts for differences in both spatial resolution and the underlying prevalence of HIV. Within the AHRI study area, prevalence is estimated at the grid-cell level (100 × 100 meters; see Section~\ref{sec:localhiv}), with a median value of 31.56 (Q1 = 30.00; Q3 - 33.16). Outside the AHRI study area, HIV prevalence estimates were available only at the district level, with a median of 26.42 (Q1 = 20.10; Q3 = 28.44) \citep{ihme-2019,ihme-2021}. Most study participants (84.3\%; 172 out of 204) have their home activity spaces located within the study area. This represents 84.9\% of men and 83.9\% of women. On the other hand, 15.7\% (32 participants) reside outside the study area. Separating participants based on the location of their residence (within or outside the AHRI study area) does not diminish the interpretability of the overall contextual exposure ($\mathcal{E}_{\text{overall}}$). Participants residing within the study area spend most of their time there (median proportion 0.96, Q1 = 0.88, Q3 = 0.99). In contrast, those living outside the study area spend significantly less time in it (median 0.33, Q1 = 0.17, Q3 = 0.42). 

\begin{figure}[htbp]
\centering
\includegraphics[width=1\textwidth]{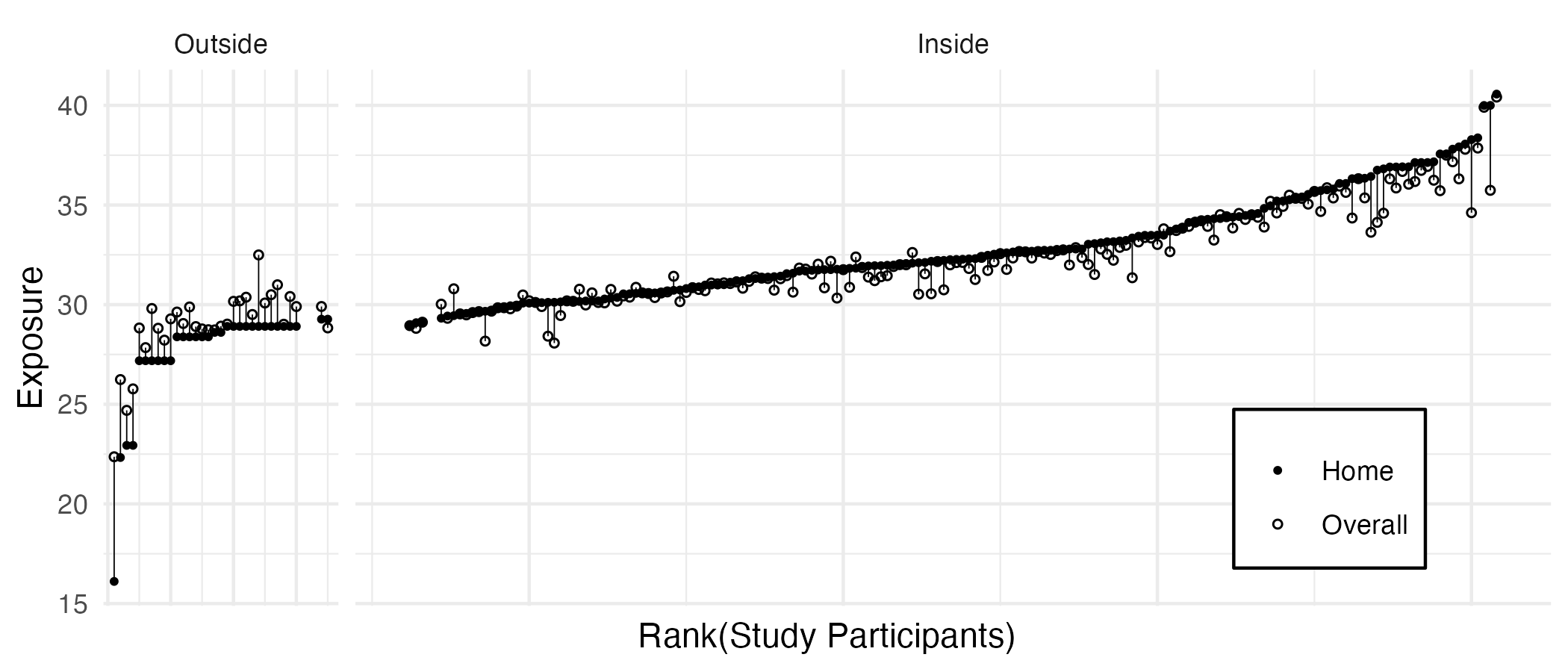}
\caption{Pairwise comparison of $\mathcal{E}_{\text{home}}$ and $\mathcal{E}_{\text{overall}}$ for all study participants. The left panel shows participants whose home activity spaces are located outside the AHRI study area, and the right panel shows those whose home activity spaces are located inside the study area. Exposure is expressed as a percentage.}
\label{fig:ttest_home_overall}
\end{figure}

We compare the HIV exposure of study participants based on their' home activity spaces ($\mathcal{E}_{\text{home}}$) with their overall HIV exposure based on  ($\mathcal{E}_{\text{overall}}$), stratified by whether their home activity spaces are located inside or outside the AHRI study area—see Figure \ref{fig:ttest_home_overall}. Participants whose home activity spaces are located inside the study area tend to have higher $\mathcal{E}_{\text{home}}$ than $\mathcal{E}_{\text{overall}}$, while the opposite pattern is observed for participants living outside the study area. Paired two-sided $t$-tests confirm these differences, with both groups showing statistically significant results ($p < 0.001$) and mean differences of $0.385$ (inside) and $-1.422$ (outside). 

Next, we exemplify how to integrate measures of HIV contextual exposure and mobility into a unified framework that identifies study participants that are at a higher risk of HIV acquisition. We chose the contextual exposure within the AHRI study area $\mathcal{E}_{\text{in}}$ as a measure that effectively reflects the level of HIV exposure among participants whose home activity spaces are located within the study area, with higher
$\mathcal{E}_{\text{in}}$ indicating a greater contextual risk. The second measure we selected is the proportion of time spent outside the AHRI study area. Prior work in the same rural setting shows that spending extended periods outside the study area—thereby separating individuals from their families and social groups—raises the risk of HIV acquisition. This is due to a higher likelihood of these individuals connecting to a new sexual network \citep{dobra-et-2017}. This pair of measures indicates a higher likelihood of HIV acquisition when a study participant is inside the AHRI study area through $\mathcal{E}_{\text{in}}$ (a local mobility measure), as well as outside the AHRI study area based on the proportion of time spent outside (an external mobility measure). Study participants are subsequently categorized into four groups: low risk (below the 40th percentile of both $\mathcal{E}_{\text{in}}$ and time outside), high risk (above the 60th percentile of both), high risk (local) (above the 60th percentile of $\mathcal{E}_{\text{in}}$ only), and high risk (external) (above the 60th percentile of time outside only).
 
\begin{figure}[htbp]
\centering
\includegraphics[width=1\textwidth]{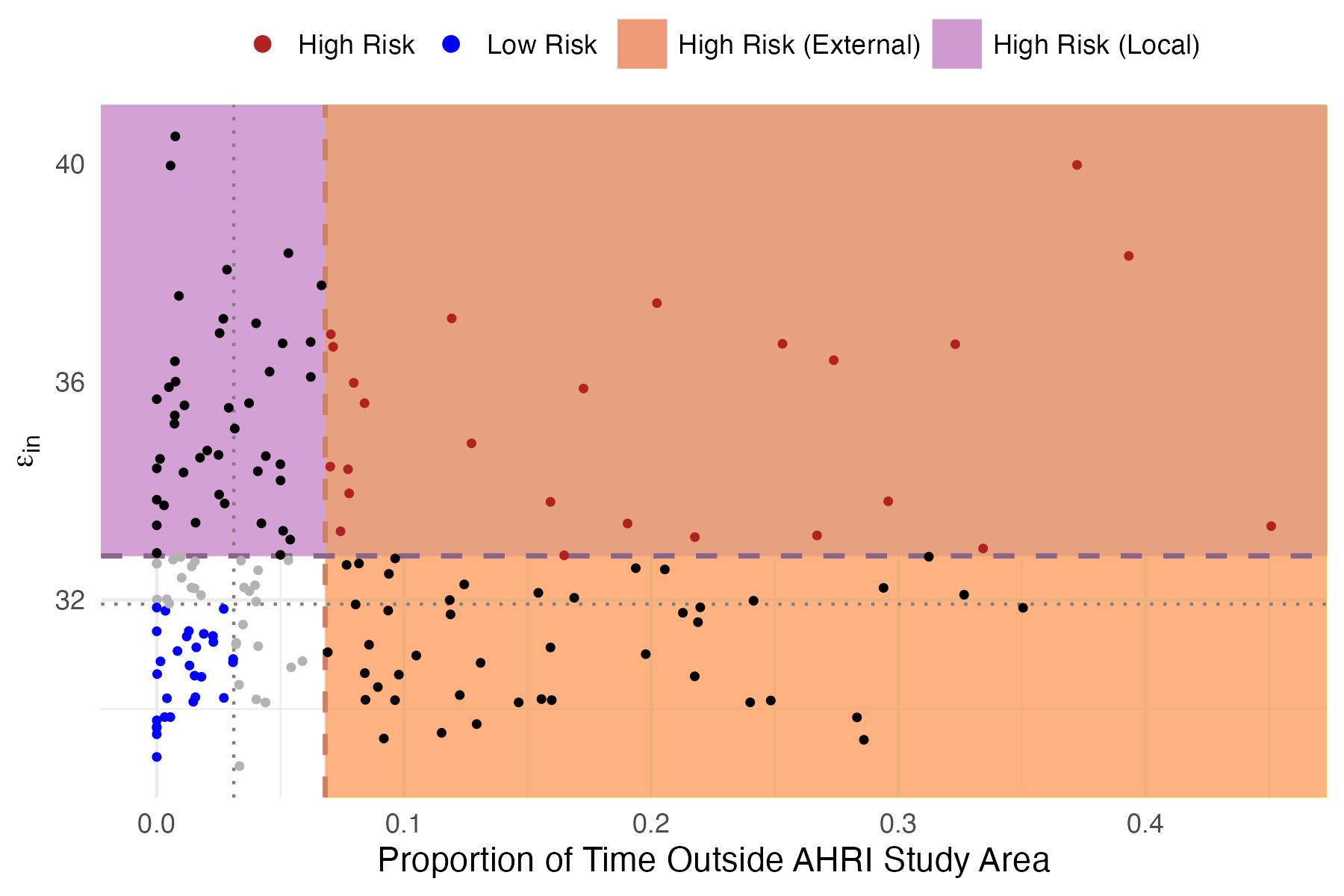}
\caption{Scatterplot of participants’ HIV risk profiles based on time spent outside the AHRI study area and $\mathcal{E}_{\text{in}}$, expressed as a percentage. Vertical and horizontal lines indicate the 40th (dotted) and 60th (dashed) percentiles of each variable. Shaded areas highlight regions of high time outside (right) and high $\mathcal{E}_{\text{in}}$ (top). Points are categorized into low risk (blue), intermediate risk (black) and high risk (red) groups. Gray dots indicate study participants that are between the 40th and the 60th percentiles with respect to both variables. These are participants that were not categorized in any risk group.}
\label{fig:risk_groups}
\end{figure}

Figure~\ref{fig:risk_groups} illustrates the joint distribution of these two risk measures. Each point represents an individual participant, positioned according to their contextual exposure within the AHRI study area (vertical axis) and the proportion of time spent outside the study area (horizontal axis). The shaded regions indicate zones of elevated risk: the upper region corresponds to participants exposed to a higher HIV prevalence within the study area, while the right region identifies those who spend a larger share of their time outside the study area. Participants located in the upper-right quadrant belong to the high-risk group (red points); they experience both high internal HIV exposure and prolonged periods outside the AHRI study area. Participants located in the lower-left quadrant belong to the low-risk group (blue points); they exhibit both low internal HIV exposure and spend less time outside the AHRI study area. The study participants who fall into the other two quadrants form intermediate risk groups characterized either by high local HIV exposure $\mathcal{E}_{\text{in}}$ or by prolonged periods outside the study area. Overall, this bivariate pattern highlights how mobility amplifies or mitigates contextual HIV risk by linking the intensity of exposure to the spatial range of behavioral movement.

Table~\ref{tab:risk_group} presents demographic characteristics and summary statistics for the four risk profile groups. Study participants who belong to the low-risk group are more likely to be women than men, are slightly older compared to the average age of study participants, and tend to spend a larger proportion of their time within their home activity spaces. These findings align with broader social and economic dynamics in South Africa, where women are more likely to participate in household-based activities and spend more time at or near home \citep{StatsSA2013}. If their home locations correspond to areas of lower HIV prevalence, these individuals are more likely to fall into the low-risk category. The association between greater home-centered activity and lower HIV contextual exposure suggests that the spatial concentration of daily movements may reduce overall exposure opportunities. In contrast, study participants who belong to the high-risk group are more likely to be men than women. Participants in this group spend extended periods outside the AHRI study area, but they also devote considerable time within their home activity spaces. This indicates that, although they travel frequently, they continue to maintain strong residential connections and return home regularly. On average, they visit fewer grid cells within the study area, likely because a significant portion of their movement occurs outside its boundaries.

\begin{figure}[htbp]
\centering
\includegraphics[width=1\textwidth]{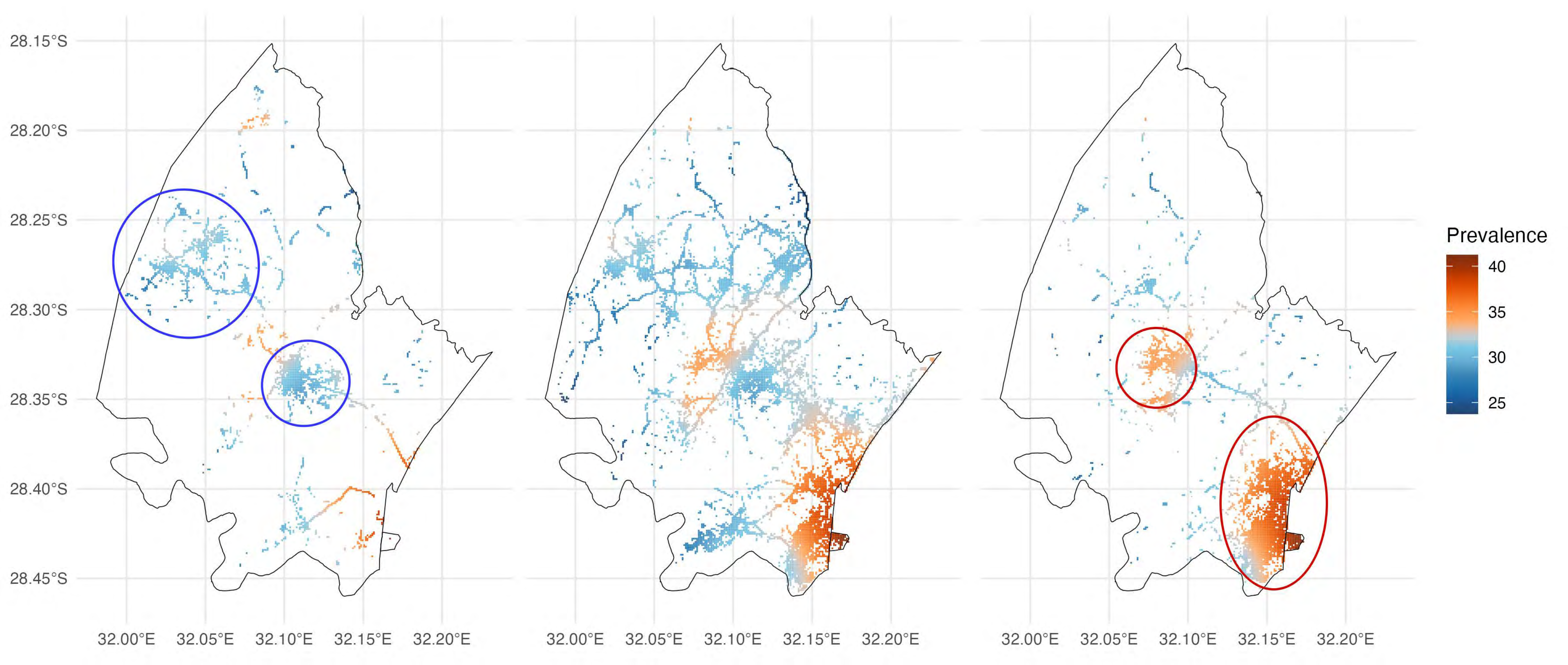}
\caption{Collective activity spaces of participants whose home activity spaces are located inside the AHRI study area, divided into four groups based on two risk factors: low-risk (left panel), high-risk (external) (middle panel) and high-risk (local) (right panel). Prevalence is expressed as a percentage.}
\label{fig:map_prev_risk}
\end{figure}

Figure~\ref{fig:map_prev_risk} displays the collective activity spaces of three out of the four risk groups. Since the high-risk group is a subset of both the high-risk (local) and high-risk (external) groups, its collective activity space is presented in Appendix \ref{app:highrisk}. The high-risk (external) group exhibits a broader spatial extent than the other groups; however, as shown in Table~\ref{tab:risk_group}, the average number of grid cells visited per person is smaller. This indicates that individuals in this group tend to travel to distinct locations that are not commonly visited by others. In other words, their individual activity spaces are relatively small, but their combined collective footprint is extensive. This pattern is consistent with the fact that these participants spend more time outside the AHRI study area, which encompasses a much larger region and comprises more spatially dispersed destinations.

Human mobility among all high-risk groups is concentrated in the central and southeastern portions of the study area, which correspond to the high-prevalence zones identified in Figure~\ref{fig:hivprevalence}. Participants in the low-risk group exhibit cluster movement in the central and northwest regions, where prevalence levels are lower. Although the movement patterns of some study participants in this group extend into areas of higher-prevalence, their overall exposure remains limited because, as shown in Table~\ref{tab:risk_group}, they predominantly spend their time within their home activity spaces. Participants whose overall exposure differs substantially from their home exposure are likely to experience contextual environments that differ markedly from their residential surroundings. Examining these individuals provides information on how mobility contributes to missed or underestimated exposure risks.

We develop a clustering approach for study participants that captures how HIV contextual exposure changes as the activity space expands beyond the home activity space. The key objective is to determine whether the study participants encounter areas of higher or lower HIV prevalence as they travel beyond their most visited locations. Since home location represents the baseline exposure context of each participant, we compare mobility-based exposure measures against this home-based reference. By doing so, we can control for the influence of daily movement on HIV exposure, while also accounting for differences in residential HIV prevalence levels.

\begin{figure}[htbp]
\centering
\includegraphics[width=1\textwidth]{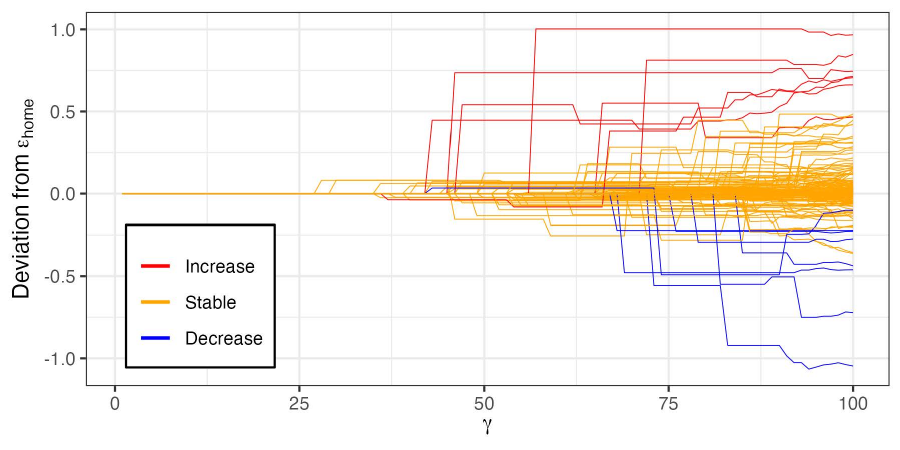}
\caption{Clustering of study participants with home activity spaces inside the AHRI study area into 3 groups based on deviations of contextual exposure from the home activity space HIV exposure across activity space levels 50–95.}
\label{fig:cls_exp}
\end{figure}

To this end, we applied $k$-means clustering based on the difference between $\mathcal{E}_{\text{home}}$ and the contextual exposure associated with $\mathcal{AS}_\gamma$ for $\gamma \in [50, 95]$. For each study participant and for each activity-space level, we compute the contextual exposure to HIV as a weighted average of the cell-level estimates of HIV prevalence on the grid in cells $\mathcal{AS}_\gamma$. We then subtract $\mathcal{E}_{\text{home}}$ to determine whether expanding one’s routine beyond home shifts exposure to higher or lower-prevalence contexts. The study participants were then categorized into three groups: increase ($\mathcal{E}_{\text{home}} < \mathcal{E}_{\text{overall}}$), stable ($\mathcal{E}_{\text{home}} \approx \mathcal{E}_{\text{overall}}$), and decrease ($\mathcal{E}_{\text{home}} > \mathcal{E}_{\text{overall}}$), as shown in Figure~\ref{fig:cls_exp}.

\begin{figure}[htbp]
\centering
\includegraphics[width=1\textwidth]{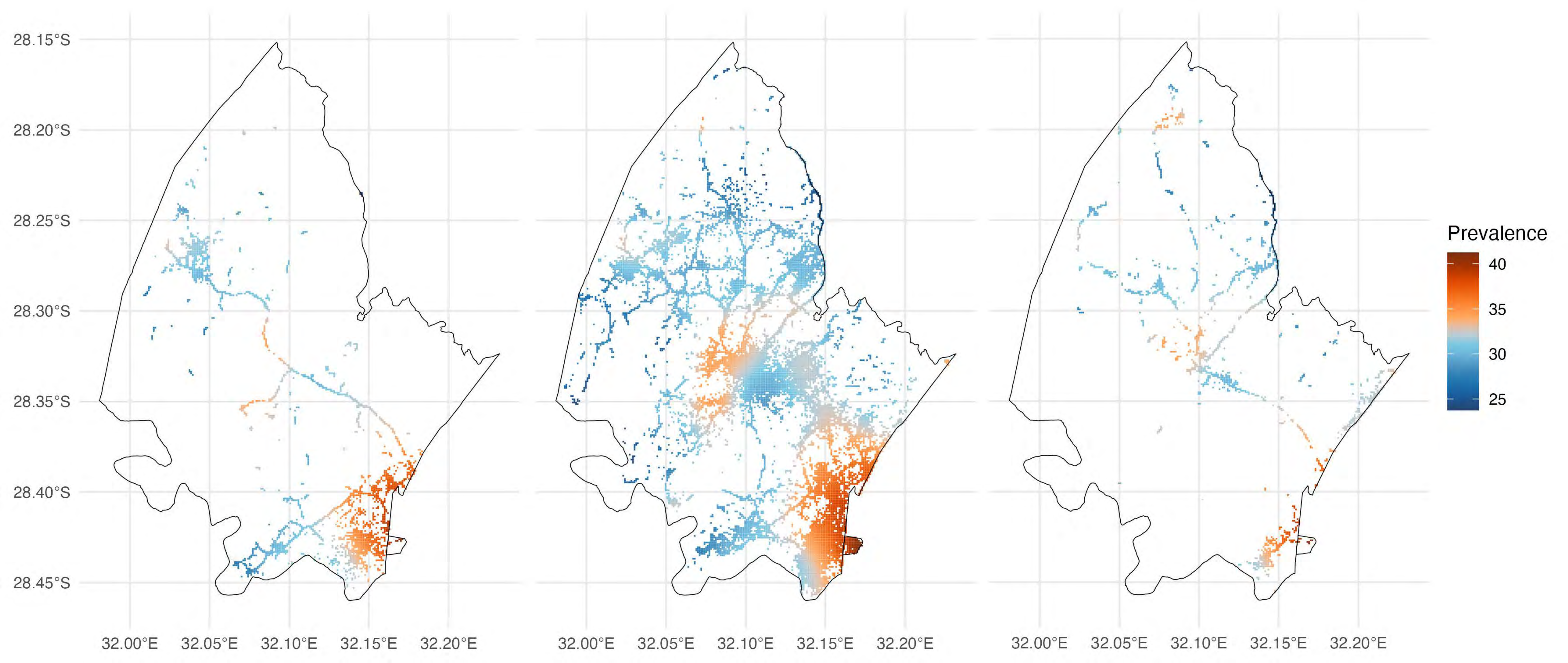}
\caption{Collective 
activity spaces for study participants in exposure clusters based on deviations of contextual exposure from the home activity space exposure across activity space levels 50–95: increase (left panel), stable (middle panel), decrease (right panel). Prevalence is expressed as a percentage.}
\label{fig:map_cls_exp}
\end{figure}

The collective activity spaces of individuals within each cluster are illustrated in Figure~\ref{fig:map_cls_exp}. Participants in the increase cluster primarily reside in areas of low prevalence (blue circles), but travel toward the southeastern, central, and northern parts of the study area, where the prevalence of HIV is higher. In contrast, participants in the decrease cluster are mainly located in areas of high prevalence (red circles) and tend to move toward northern regions, where prevalence levels are lower. The majority of the participants belong to the stable group, whose collective activity space closely mirrors that of the general study population, indicating consistent exposure at different levels of the activity space.

In Table~\ref{tab:risk_group}, the majority of the participants belong to the stable exposure group, indicating that their contextual exposure to HIV remains relatively constant as their activity spaces expand beyond their home area. Study participants in the increase and decrease clusters are fewer, but display distinct behavioral and spatial characteristics. Those in the increase group tend to spend less time within their home activity spaces, are more often men, and exhibit greater mobility at the 95\% activity space level ($\mathcal{AS}_{95}$). In contrast, participants in the decrease group are more frequently women and have larger total activity spaces at the 100\% level ($\mathcal{AS}_{100}$), while their $\mathcal{AS}_{95}$ remains relatively compact. This suggests that individuals in this group spend most of their time in fewer locations but occasionally make one-time or infrequent trips to more distant regions, thereby expanding their overall activity space without increasing routine mobility. Because the increase and decrease groups contain relatively few participants, these patterns should be interpreted with caution and viewed as suggestive rather than conclusive.

\section{Conclusions} \label{sec:conclusions}

Our methodological contributions go beyond the particular context of a GPS mobility study designed for an HIV surveillance site. Our techniques for contextual exposure estimation based on activity spaces can be applied to GPS mobility studies aimed at determining the sociodemographic and behavioral characteristics of individuals who are at an increased risk of acquiring other infectious diseases, or to assess dynamic exposure to environmental hazards or social environments. To the best of our knowledge, determining groups of study participants with specific risk profiles by varying the levels of GPS-based activity spaces is novel and represents a generalization of previous approaches to activity space estimation that view activity spaces as fixed spatial areas \--- see, for example, \citet{chen-dobra-2020} and the references therein. In our framework, activity spaces are functions that map levels into spatial areas that can grow or remain constant as the level increases. This provides a more flexible conceptual view of activity spaces, which can be further applied in the development of related modeling frameworks for representing human mobility.

Our empirical findings are also very interesting. Based on coarser mobility data such as residential locations, recent studies have determined that the mobility patterns of men and women living in South Africa and other sub-Saharan countries are quite similar \--- see, for example, \citet{dobra-et-2017} and the references therein. However, the GPS data collected in the Sesikhona study, due to their much higher spatial and temporal resolution, provide evidence that men and women engage in different patterns of movement in their activities of daily living. These differential spatiotemporal trajectories are associated with varying levels of risk for HIV acquisition, which we characterized through the measures we discussed. 

We also highlight our results that contrast the assessment of contextual exposure at home locations with that in activity spaces of varying levels. Aggregating exposure across all locations visited by study participants, weighted by the proportion of time spent at these locations, captures the full spectrum of environments encountered during their daily activities. In principle, such an overall exposure measure should always provide a closer representation of true contextual exposure than traditional home-based estimates. However, its practical advantage depends on whether overall exposure meaningfully differs from home exposure, as large discrepancies would indicate that relying solely on residential location omits substantial variation in exposure risk. The conventional home-based approach implicitly assumes that exposure is limited to the immediate residential area, thereby overlooking mobility and interactions with spatially heterogeneous risk environments. In contrast, GPS-derived measures reflect both where and for how long individuals are exposed, providing a more comprehensive view of possible high-risk health contexts.

\section*{Funding}

This research was supported in part by the National Institutes of Health (NIH) grant (1R01MH131480-01). The Sesikhona GPS study was funded by the German Research Foundation (BA2067/14-1). The AHRI demographic surveillance information system and population intervention program was funded by the Wellcome Trust (227167/A/23/Z) and the South Africa Population Research Infrastructure Network (funded by the South African Department of Science and Technology and hosted by the South African Medical Research Council). The work of KHT was supported by the NIH/NIAID (K01AI193303) and by a 2024 New Investigator Award from the University of Washington / Fred Hutch Center for AIDS Research, an NIH-funded program under award number AI027757. The findings and conclusions in this paper are those of the author(s) and do not necessarily represent the official position of the funding source. The funders had no role in the study design, data collection and analysis, decision to publish or preparation of the manuscript.

%\bibliographystyle{plainnat}
%\bibliography{references}

\begin{thebibliography}{38}
\providecommand{\natexlab}[1]{#1}
\providecommand{\url}[1]{\texttt{#1}}
\expandafter\ifx\csname urlstyle\endcsname\relax
  \providecommand{\doi}[1]{doi: #1}\else
  \providecommand{\doi}{doi: \begingroup \urlstyle{rm}\Url}\fi

\bibitem[Birdthistle et~al.(2019)Birdthistle, Tanton, Tomita, de~{G}raaf, Schaffnit, Tanser, and Slaymaker]{birdthistle-et-2019}
I.~Birdthistle, C.~Tanton, A.~Tomita, K.~de~{G}raaf, S.~B. Schaffnit, F.~Tanser, and E.~Slaymaker.
\newblock Recent levels and trends in {H}{I}{V} incidence rates among adolescent girls and young women in ten high-prevalence {A}frican countries: a systematic review and meta-analysis.
\newblock \emph{Lancet Glob Health}, 7:\penalty0 e1521--e1540, 2019.

\bibitem[Bulstra et~al.(2020)Bulstra, Hontelez, Giardina, Steen, Nagelkerke, B\"{a}rnighausen, and de~Vlas]{bulstra-et-2020}
Caroline~A. Bulstra, Jan A.~C. Hontelez, Federica Giardina, Richard Steen, Nico J.~D. Nagelkerke, Till B\"{a}rnighausen, and Sake~J. de~Vlas.
\newblock Mapping and characterising areas with high levels of {H}{I}{V} transmission in sub-{S}aharan {A}frica: A geospatial analysis of national survey data.
\newblock \emph{PLOS Medicine}, 17:\penalty0 1--19, 2020.

\bibitem[Byrnes et~al.(2015)Byrnes, Miller, Wiebe, Morrison, Remer, and Wiehe]{byrnes-et-2015}
H.~F. Byrnes, B.~A. Miller, D.~J. Wiebe, C.~N. Morrison, L.~G. Remer, and S.~E. Wiehe.
\newblock Tracking adolescents with global positioning system-enabled cell phones to study contextual exposures and alcohol and marijuana use: A pilot study.
\newblock \emph{The Journal of adolescent health : official publication of the Society for Adolescent Medicine}, 57:\penalty0 245--247, 2015.

\bibitem[Chaix et~al.(2013)Chaix, M\'{e}line, Duncan, Merrien, Karusisi, Perchoux, Lewin, Labadi, and Kestens]{chaix-et-2013}
Basile Chaix, Julie M\'{e}line, Scott Duncan, Claire Merrien, Noëlla Karusisi, Camille Perchoux, Antoine Lewin, Karima Labadi, and Yan Kestens.
\newblock {G}{P}{S} tracking in neighborhood and health studies: A step forward for environmental exposure assessment, a step backward for causal inference?
\newblock \emph{Health and Place}, 21:\penalty0 46--51, 2013.

\bibitem[Chen and Dobra(2020)]{chen-dobra-2020}
Y.-C. Chen and A.~Dobra.
\newblock Measuring human activity spaces from gps data with density ranking and summary curves.
\newblock \emph{Annals of Applied Statistics}, 14:\penalty0 409--432, 2020.

\bibitem[Clark et~al.(2025)Clark, Zilber, Schmitt, Fargo, Reif, Motsinger-Reif, and Messier]{clark-et-2025}
Lara~P. Clark, Daniel Zilber, Charles Schmitt, David~C. Fargo, David~M. Reif, Alison~A. Motsinger-Reif, and Kyle~P. Messier.
\newblock A review of geospatial exposure models and approaches for health data integration.
\newblock \emph{Journal of Exposure Science and Environmental Epidemiology}, 35:\penalty0 131--148, 2025.

\bibitem[Dobra et~al.(2017)Dobra, B\"{a}rnighausen, Vandormael, and Tanser]{dobra-et-2017}
A.~Dobra, T.~B\"{a}rnighausen, A.~Vandormael, and F.~Tanser.
\newblock Space-time migration patterns and risk of {H}{I}{V} acquisition in rural {S}outh {A}frica.
\newblock \emph{AIDS}, 31:\penalty0 137--145, 2017.

\bibitem[Dong et~al.(2020)Dong, Chen, and Dobra]{dong2020statistical}
Zhihang Dong, Yen-Chi Chen, and Adrian Dobra.
\newblock A statistical framework for measuring the temporal stability of human mobility patterns.
\newblock \emph{Journal of Applied Statistics}, pages 1--19, 2020.

\bibitem[Duncan et~al.(2018)Duncan, Chaix, Regan, Park, Draper, Goedel, Gipson, Guilamo-Ramos, Halkitis, Brewer, and Hickson]{duncan-et-2018}
Dustin~T. Duncan, Basile Chaix, Seann~D. Regan, Su~Hyun Park, Cordarian Draper, William~C. Goedel, June~A. Gipson, Vincent Guilamo-Ramos, Perry~N. Halkitis, Russell Brewer, and De{M}arc~A. Hickson.
\newblock Collecting mobility data with {G}{P}{S} methods to understand the hiv environmental riskscape among young black men who have sex with men: A multi-city feasibility study in the deep south.
\newblock \emph{AIDS and Behavior}, 22:\penalty0 3057--3070, 2018.

\bibitem[Dwyer-Lindgren et~al.(2019)Dwyer-Lindgren, Cork, Sligar, Steuben, Wilson, Provost, Mayala, VanderHeide, Collison, Hall, Biehl, Carter, Frank, Douwes-Schultz, Burstein, Casey, Deshpande, Earl, El~Bcheraoui, Farag, Henry, Kinyoki, Marczak, Nixon, Osgood-Zimmerman, Pigott, Reiner, Ross, Schaeffer, Smith, Davis~Weaver, Wiens, Eaton, Justman, Opio, Sartorius, Tanser, Wabiri, Piot, Murray, and Hay]{ihme-2019}
Laura Dwyer-Lindgren, Michael~A. Cork, Amber Sligar, Krista~M. Steuben, Kate~F. Wilson, Naomi~R. Provost, Benjamin~K. Mayala, John~D. VanderHeide, Michael~L. Collison, Jason~B. Hall, Molly~H. Biehl, Austin Carter, Tahvi Frank, Dirk Douwes-Schultz, Roy Burstein, Daniel~C. Casey, Aniruddha Deshpande, Lucas Earl, Charbel El~Bcheraoui, Tamer~H. Farag, Nathaniel~J. Henry, Damaris Kinyoki, Laurie~B. Marczak, Molly~R. Nixon, Aaron Osgood-Zimmerman, David Pigott, Robert~C. Reiner, Jennifer~M. Ross, Lauren~E. Schaeffer, David~L. Smith, Nicole Davis~Weaver, Kirsten~E. Wiens, Jeffrey~W. Eaton, Jessica~E. Justman, Alex Opio, Benn Sartorius, Frank Tanser, Njeri Wabiri, Peter Piot, Christopher J.~L. Murray, and Simon~I. Hay.
\newblock Mapping {H}{I}{V} prevalence in sub-{S}aharan {A}frica between 2000 and 2017.
\newblock \emph{Nature}, 570:\penalty0 189--193, 2019.

\bibitem[Entwisle(2007)]{entwisle-2007}
B.~Entwisle.
\newblock Putting people into place.
\newblock \emph{Demography}, 44:\penalty0 687--703, 2007.

\bibitem[Friis and Sellers(2009)]{friis-sellers-2009}
Robert~H. Friis and Thomas~A. Sellers.
\newblock \emph{Epidemiology for Public Health Practice}.
\newblock Jones and Bartlett Publishers, LLC, fourth edition, 2009.

\bibitem[Gareta et~al.(2021)Gareta, Baisley, Mngomezulu, Smit, Khoza, Nxumalo, Dreyer, Dube, Majozi, Ording-{J}esperson, Ehlers, Harling, Shahmanesh, Siedner, Hanekom, and K.]{gareta-et-2021}
D.~Gareta, K.~Baisley, T.~Mngomezulu, T.~Smit, T.~Khoza, S.~Nxumalo, J.~Dreyer, S.~Dube, N.~Majozi, G.~Ording-{J}esperson, E.~Ehlers, G.~Harling, M.~Shahmanesh, M.~Siedner, W.~Hanekom, and Herbst K.
\newblock Cohort profile update: {A}frica {C}entre {D}emographic {I}nformation {S}ystem ({A}{C}{D}{I}{S}) and population-based {H}{I}{V} survey.
\newblock \emph{International Journal of Epidemiology}, 50:\penalty0 33--34, 2021.

\bibitem[Gesler and Meade(1988)]{gesler-meade-1988}
W.~M. Gesler and M.~S. Meade.
\newblock Locational and population factors in health care-seeking behavior in {S}avannah, {G}eorgia.
\newblock \emph{Health Services Research}, 23:\penalty0 443--462, 1988.

\bibitem[Glasgow et~al.(2019)Glasgow, Le, {Scott Geller}, Fan, and Hankey]{glasgow-et-2019}
Trevin~E. Glasgow, Huyen~T.K. Le, E.~{Scott Geller}, Yingling Fan, and Steve Hankey.
\newblock How transport modes, the built and natural environments, and activities influence mood: A {G}{P}{S} smartphone app study.
\newblock \emph{Journal of Environmental Psychology}, 66:\penalty0 101345, 2019.

\bibitem[Hassani et~al.(2023)Hassani, De~Haro, Flores, Emish, Kim, Kelani, Ugarte, Hightow-Weidman, Castel, Li, Theall, and Young]{hassani-et-2023}
Maryam Hassani, Cristina De~Haro, Lidia Flores, Mohamed Emish, Seungjun Kim, Zeyad Kelani, Dominic~Arjuna Ugarte, Lisa Hightow-Weidman, Amanda Castel, Xiaoming Li, Katherine~P Theall, and Sean Young.
\newblock Exploring mobility data for enhancing hiv care engagement in {B}lack/{A}frican {A}merican and {H}ispanic/{L}atinx individuals: a longitudinal observational study protocol.
\newblock \emph{BMJ Open}, 13:\penalty0 e079900, 2023.

\bibitem[IHME(2021)]{ihme-2021}
{ } IHME.
\newblock Sub-{S}aharan {A}frica {H}{I}{V} incidence and mortality geospatial estimates 2000-2018.
\newblock Technical report, Seattle, United States of America: Institute for Health Metrics and Evaluation (IHME), 2021.

\bibitem[Joint~United~Nations~Programme~on {H}{I}{V}/{A}{I}{D}{S}(2024)]{UNAIDS2024report}
{}~Joint~United~Nations~Programme~on {H}{I}{V}/{A}{I}{D}{S}.
\newblock The {U}rgency of {N}ow: {A}{I}{D}{S} at a {C}rossroads: 2024 {U}{N}{A}{I}{D}{S} global {A}{I}{D}{S} {U}pdate.
\newblock Technical report, Unied Nations, Geneva, 2024.
\newblock URL \url{https://crossroads.unaids.org/}.
\newblock June 30, 2025.

\bibitem[Kandwal et~al.(2009)Kandwal, Garg, and Garg]{kandwal-et-2009}
Rashmi Kandwal, P.K. Garg, and R.D. Garg.
\newblock Health {G}{I}{S} and {H}{I}{V}/{A}{I}{D}{S} studies: Perspective and retrospective.
\newblock \emph{Journal of Biomedical Informatics}, 42:\penalty0 748--755, 2009.

\bibitem[Kwan(2012)]{kwan-2012}
M.-P. Kwan.
\newblock The uncertain geographic context problem.
\newblock \emph{Annals of the Association of American Geographers}, 102:\penalty0 958--968, 2012.

\bibitem[Kwan(2009)]{kwan-2009}
Mei-Po Kwan.
\newblock From place-based to people-based exposure measures.
\newblock \emph{Social Science \& Medicine}, 69:\penalty0 1311--1313, 2009.

\bibitem[Larmarange et~al.(2015)Larmarange, Mossong, B\"{a}rnighausen, and Newell]{larmarange-et-2015}
J.~Larmarange, J.~Mossong, T.~B\"{a}rnighausen, and M.~L. Newell.
\newblock Participation dynamics in population-based longitudinal {H}{I}{V} surveillance in rural {S}outh {A}frica.
\newblock \emph{PLoS One}, 10:\penalty0 e0123345, 2015.

\bibitem[Marquet et~al.(2022)Marquet, Hirsch, Kerr, Jankowska, Mitchell, Hart, Laden, Hipp, and James]{marquet-et-2022}
Oriol Marquet, Jana~A. Hirsch, Jacqueline Kerr, Marta~M. Jankowska, Jonathan Mitchell, Jaime~E. Hart, Francine Laden, J.~Aaron Hipp, and Peter James.
\newblock {G}{P}{S}-based activity space exposure to greenness and walkability is associated with increased accelerometer-based physical activity.
\newblock \emph{Environment International}, 165:\penalty0 107317, 2022.

\bibitem[Marquet et~al.(2023)Marquet, Tello-Barsocchini, Couto-Trigo, G\'{o}mez-Varo, and Maciejewska]{marquet-et-2023}
Oriol Marquet, Jose Tello-Barsocchini, Daniel Couto-Trigo, Irene G\'{o}mez-Varo, and Monika Maciejewska.
\newblock Comparison of static and dynamic exposures to air pollution, noise, and greenness among seniors living in compact-city environments.
\newblock \emph{International Journal of Health Geographics}, 22:\penalty0 1--16, 2023.

\bibitem[Mathenjwa et~al.(2025)Mathenjwa, Okango, Tram, Inghels, Cuadros, Kim, Walsh, B\"{a}rnighausen, Dobra, and Tanser]{mathenjwa-et-2025}
T.~Mathenjwa, E.~Okango, K.~H. Tram, M.~Inghels, D.~Cuadros, H.~Y. Kim, F.~Walsh, T.~B\"{a}rnighausen, A.~Dobra, and F.~Tanser.
\newblock Leveraging smartphone mobility data to understand {H}{I}{V} risk among rural {S}outh {A}frican young adults: a pilot study.
\newblock \emph{JMIR mHealth and uHealth}, 13:\penalty0 e67519, 2025.

\bibitem[Skar et~al.(2013)Skar, Albert, and Leitner]{skar-et-2013}
H.~Skar, J.~Albert, and T.~Leitner.
\newblock Towards estimation of {H}{I}{V}-1 date of infection: a time-continuous {I}g{G}-model shows that seroconversion does not occur at the midpoint between negative and positive tests.
\newblock \emph{PloS One}, 8:\penalty0 e60906, 2013.

\bibitem[{Statistics South Africa}(2013)]{StatsSA2013}
{Statistics South Africa}.
\newblock {A Survey of Time Use, 2010}, 2013.

\bibitem[Tanser et~al.(2008)Tanser, Hosegood, B\"{a}rnighausen, Herbst, Nyirenda, Muhwava, Newell, Viljoen, Mutevedzi, and Newell]{tanser-et-2008}
F.~Tanser, V.~Hosegood, T.~B\"{a}rnighausen, K.~Herbst, M.~Nyirenda, W.~Muhwava, C.~Newell, J.~Viljoen, T.~Mutevedzi, and M.~L. Newell.
\newblock Cohort profile: {A}frica {C}entre {D}emographic {I}nformation {S}ystem ({A}{C}{D}{I}{S}) and population-based {H}{I}{V} survey.
\newblock \emph{International Journal of Epidemiology}, 37:\penalty0 956--962, 2008.

\bibitem[Tanser et~al.(2009)Tanser, B\"{a}rnighausen, Cooke, and Newell]{tanser-et-2009}
F.~Tanser, T.~B\"{a}rnighausen, G.~S. Cooke, and M.-L. Newell.
\newblock Localized spatial clustering of {H}{I}{V} infections in a widely disseminated rural {S}outh {A}frican epidemic.
\newblock \emph{International Journal of Epidemiology}, 38:\penalty0 1008--1016, 2009.

\bibitem[Tanser et~al.(2015)Tanser, B\"{a}rnighausen, Vandormael, and Dobra]{tanser-et-2015}
F.~Tanser, T.~B\"{a}rnighausen, A.~Vandormael, and A.~Dobra.
\newblock {H}{I}{V} treatment cascade in migrants and mobile populations.
\newblock \emph{Current opinion in {H}{I}{V} and {A}{I}{D}{S}}, 10:\penalty0 430--438, 2015.

\bibitem[Vandormael et~al.(2018)Vandormael, Dobra, B\"{a}rnighausen, de~{O}liveira, and Tanser]{vandormael-et-2018}
A.~Vandormael, A.~Dobra, T.~B\"{a}rnighausen, T.~de~{O}liveira, and F.~Tanser.
\newblock Incidence rate estimation, periodic testing and the limitations of the mid-point imputation approach.
\newblock \emph{International Journal of Epidemiology}, 47:\penalty0 236--245, 2018.

\bibitem[Vandormael et~al.(2020)Vandormael, Tanser, Cuadros, and Dobra]{vandormael-et-2020}
A.~Vandormael, F.~Tanser, D.~Cuadros, and A.~Dobra.
\newblock Estimating trends in the incidence rate with interval censored data and time-dependent covariates.
\newblock \emph{Statistical Methods in Medical Research}, 29:\penalty0 272--281, 2020.

\bibitem[Waller and Gotway(2004)]{waller-gotway-2004}
L.~A. Waller and C.~A. Gotway.
\newblock \emph{Applied Spatial Statistics for Public Health Data}, volume 368.
\newblock John Wiley \& Sons, 2004.

\bibitem[Wei et~al.(2025)Wei, Helbich, Fl\:{u}ckiger, Shen, Vlaanderen, Jeong, Probst-Hensch, {de Hoogh}, Hoek, and Vermeulen]{wei-et-2025}
Lai Wei, Marco Helbich, Benjamin Fl\:{u}ckiger, Youchen Shen, Jelle Vlaanderen, Ayoung Jeong, Nicole Probst-Hensch, Kees {de Hoogh}, Gerard Hoek, and Roel Vermeulen.
\newblock Variability in mobility-based air pollution exposure assessment: Effects of {G}{P}{S} tracking duration and temporal resolution of air pollution maps.
\newblock \emph{Environment International}, 198:\penalty0 109454, 2025.

\bibitem[Wiehe et~al.(2008)Wiehe, Carroll, Liu, Hoch, Jeffery S~Wilson, and Fortenberry]{wiehe-et-2008}
S.~E. Wiehe, A.~E. Carroll, K.~L. Liu, G.~C.~Haberkorn, S.~C. Hoch, J.~S. Jeffery S~Wilson, and J.~D. Fortenberry.
\newblock Using {G}{P}{S}-enabled cell phones to track the travel patterns of adolescents.
\newblock \emph{International Journal of Health Geographics}, 7:\penalty0 1--11, 2008.

\bibitem[Wong et~al.(2021)Wong, Olivier, Gunda, Koole, Surujdeen, Gareta, Munatsi, Modise, Dreyer, Nxumalo, Smit, Ording-{J}espersen, Mpofana, Khan, Sikhosana, Moodley, Shen, Khoza, Mhlongo, Bucibo, Nyamande, Baisley, Cuadros, Tanser, Grant, Herbst, Seeley, Hanekom, Ndung\'u, Siedner, and Pillay]{wong-et-2021}
E.B. Wong, S.~Olivier, R.~Gunda, O.~Koole, A.~Surujdeen, D.~Gareta, D.~Munatsi, T.~H. Modise, J.~Dreyer, S.~Nxumalo, T.~K. Smit, G.~Ording-{J}espersen, I.~B. Mpofana, K.~Khan, Z.~E.~L. Sikhosana, S.~Moodley, Y.~J. Shen, T.~Khoza, N.~Mhlongo, S.~Bucibo, K.~Nyamande, K.~J. Baisley, D.~Cuadros, F.~Tanser, A.~D. Grant, K.~Herbst, J.~Seeley, W.~A. Hanekom, T.~Ndung\'u, M.~J. Siedner, and D.~Pillay.
\newblock Convergence of infectious and non-communicable disease epidemics in rural {S}outh {A}frica: a cross-sectional, population-based multimorbidity study.
\newblock \emph{Lancet Glob Health}, 9:\penalty0 e967--e976, 2021.

\bibitem[Yi et~al.(2019)Yi, Wilson, Mason, Habre, Wang, and Dunton]{yi-et-2019}
L.~Yi, J.~P. Wilson, T.B. Mason, R.~Habre, S.~Wang, and G.F. Dunton.
\newblock Methodologies for assessing contextual exposure to the built environment in physical activity studies: A systematic review.
\newblock \emph{Health and Place}, 60:\penalty0 102226, 2019.

\bibitem[Zuo et~al.(2016)Zuo, Xia, Liu, and Qiao]{zuo-et-2016}
J.~Zuo, H.~Xia, S.~Liu, and Y.~Qiao.
\newblock Mapping urban environmental noise using smartphones.
\newblock \emph{Sensors}, 16:\penalty0 1692, 2016.

\end{thebibliography}

\begin{landscape}
\begin{table}[htbp]
\centering
{\begin{tabular}{llccccc}
\hline
 & \textbf{Group} & $n$ (M/F\%)\textsuperscript{a} & Age\textsuperscript{b} & \%Time\textsuperscript{c} & $|\mathcal{AS}(\mathcal{U}_{\text{in}})|$\textsuperscript{d} & $|\mathcal{AS}_{95}(\mathcal{U}_{\text{in}})|$\textsuperscript{e} \\
\hline
\multirow{4}{*}{Risk Profiles} 
 & Low Risk              & 59(21.9/32.8) & 27.1 $\pm$ 3.3 & 73.1 $\pm$ 18.1 & 158.6 $\pm$ 149.6 & 4.1 $\pm$ 4.3 \\
 & High Risk             & 25(15.1/10.7) & 25.8 $\pm$ 3.1 & 58.3 $\pm$ 15.6 & 152.3 $\pm$ 137.2 & 5.6 $\pm$ 4.6 \\
 & High Risk (Local)     & 44(20.5/22.1) & 26.3 $\pm$ 3.5 & 62.1 $\pm$ 15.6 & 264.3 $\pm$ 203.6 & 5.8 $\pm$ 4.8 \\
 & High Risk (External)  & 44(27.4/18.3) & 26.0 $\pm$ 3.0 & 69.6 $\pm$ 19.7 & 114.5 $\pm$ 118.0 & 4.7 $\pm$ 3.2 \\
\hline
\multirow{3}{*}{Exposure Clusters}
 & Increase              & 7(5.5/2.3)    & 25.1 $\pm$ 3.1 & 48.5 $\pm$ 11.9 & 214.3 $\pm$ 175.1 & 9.9 $\pm$ 6.8 \\
 & Stable                & 155(76.7/75.6)& 26.5 $\pm$ 3.2 & 68.4 $\pm$ 18.4 & 166.5 $\pm$ 160.3 & 4.6 $\pm$ 4.1 \\
 & Decrease              & 10(2.7/6.1)   & 25.6 $\pm$ 3.9 & 62.5 $\pm$ 14.5 & 253.4 $\pm$ 223.4 & 5.4 $\pm$ 2.8 \\
\hline
\end{tabular}}
\caption{Summary statistics of participants grouped by 1) risk profiles determined by the proportion of time spent outside the AHRI study area and $\mathcal{E}_{\text{in}}$ \--- see Figure~\ref{fig:risk_groups}; and 2) clustering of exposure \--- see Figure~\ref{fig:cls_exp}. \textsuperscript{a}Number of participants in each group, with the percentages of men and women shown in parentheses, calculated relative to the total numbers of men and women in the dataset. \textsuperscript{b,c,d,e}Mean $\pm$ standard deviation. \textsuperscript{c}Proportion of time spent within the home activity space. \textsuperscript{d}Number of grid cells visited within the AHRI study area. \textsuperscript{e}Number of grid cells where the participant spent 95\% of their time within the AHRI study area.}
\label{tab:risk_group}
\end{table}
\end{landscape}

\section*{Appendix}

\subsection{Activity distribution outside the AHRI study area}\label{app:time_outside}

\begin{figure}[htbp]
\centering
\includegraphics[width=1\textwidth]{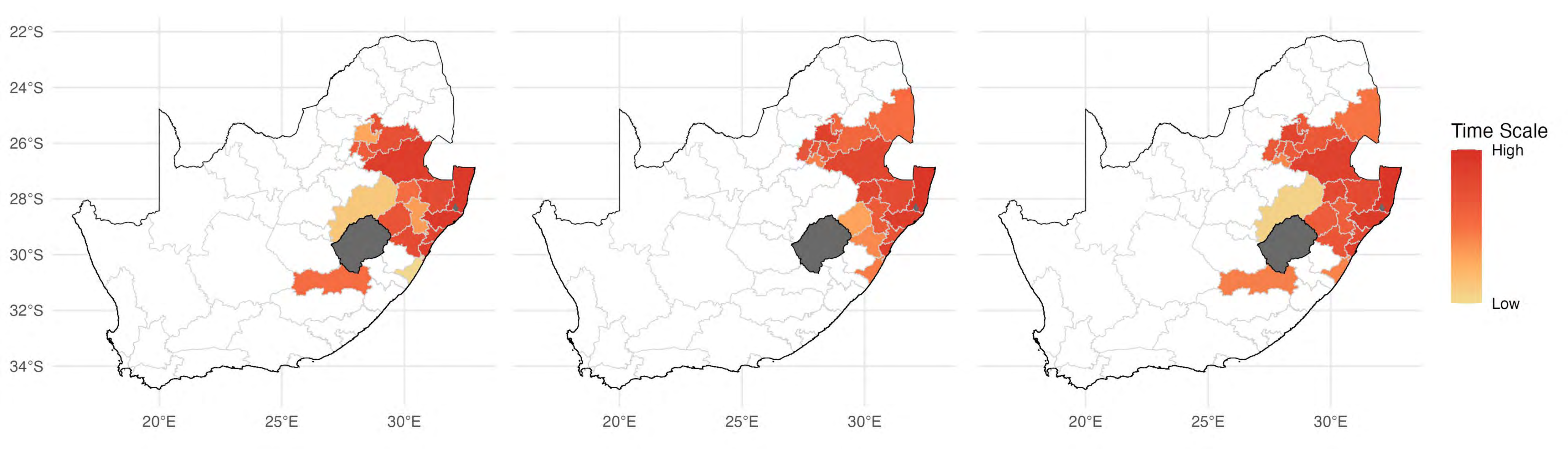}
\caption{Spatial distribution of time spent in each district of South Africa (excluding the AHRI study area) for men (left panel), women (middle panel), and all  study participants (right panel). White areas indicate no recorded time spent, while grey regions within the national boundary correspond to locations outside South Africa or the AHRI study area.} 
\label{fig:time_outside}
\end{figure}

To complement the within-area analyses (Figure \ref{fig:time_inside}) and assess whether sex differences persist beyond the study boundary, we also examine mobility outside the AHRI study area. Rather than measuring the time spent in individual grid cells as in \eqref{eq:conservativeactivity}, we now aggregate the time spent in each district outside the AHRI boundary to estimate the distribution of activity. We apply the same processing pipeline as shown in Figure \ref{fig:time_inside}, with the only difference being that grid cells are replaced by districts.
As shown in Figure \ref{fig:time_outside}, participants frequently traveled to districts adjacent to the AHRI study area, with both men and women exhibiting broadly similar patterns. Consistent with Table \ref{tab:gpssummary}, both groups spent a similar amount of time outside the study area. While strong contrasts are observed within the AHRI area, the similarity outside the study boundary may reflect the coarser spatial resolution of districts compared to grid cells, as well as the relatively limited time that participants spent outside the study area. 

\subsection{Collective activity space of the high-risk group}\label{app:highrisk}

\begin{figure}[htbp]
\centering
\includegraphics[width=\textwidth]{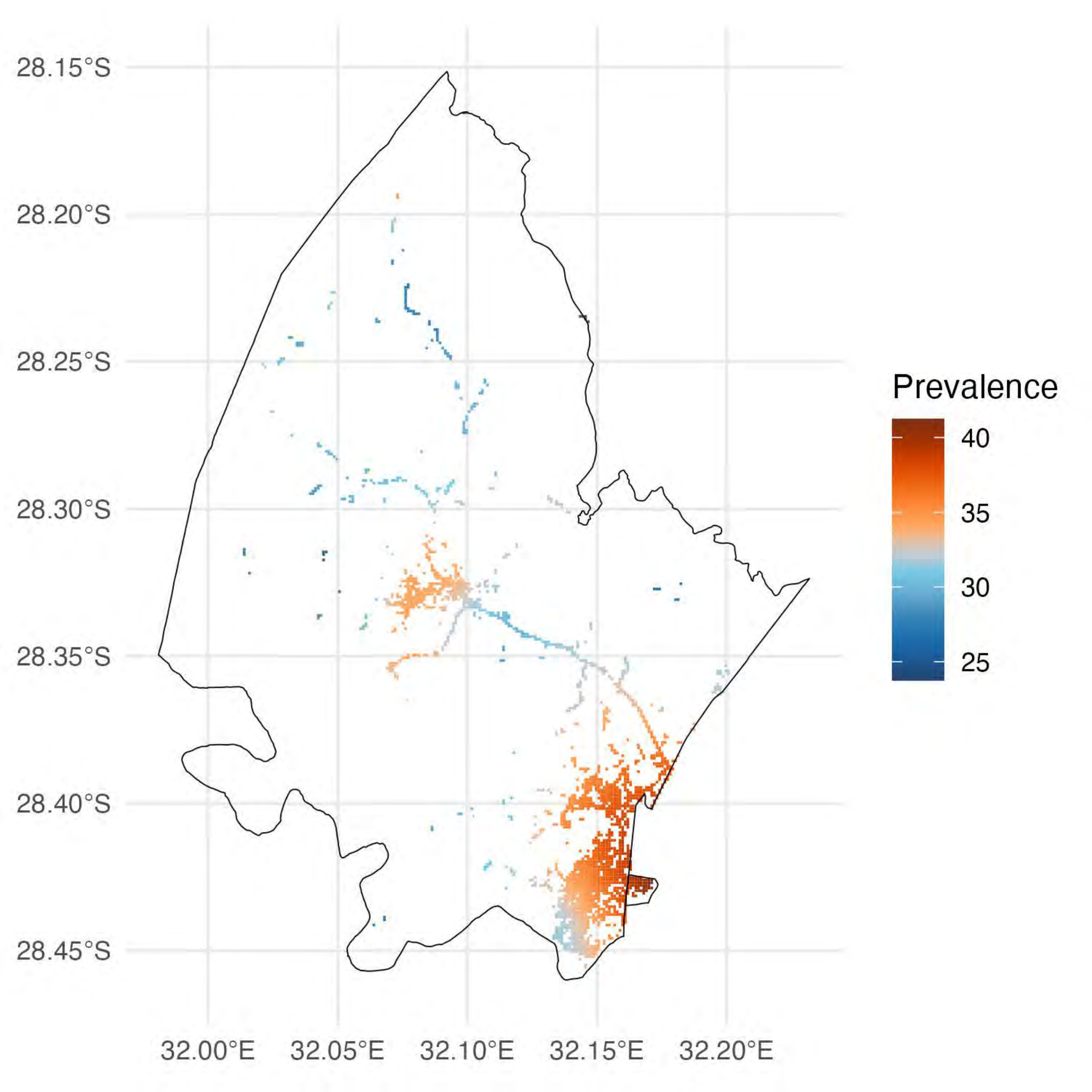}
\caption{Collective activity space of participants belonging to the high-risk group. See also Figure \ref{fig:risk_groups}. Prevalence is expressed as a percentage.}
\label{fig:map_prev_highrisk}
\end{figure}

Figure \ref{fig:risk_groups} categorizes the study participants into four distinct groups based on a local risk factor ($\mathcal{E}_\text{in}$) and an external risk factor, specifically the proportion of time spent outside the AHRI study area. The collective activity space for participants in the high-risk group is illustrated in Figure \ref{fig:map_prev_highrisk}.

\end{document}